\documentclass[12pt,twocolumn]{article}
\usepackage{amssymb}
\usepackage{amsfonts}
\usepackage{amsmath}
\usepackage{eurosym}
\usepackage{makeidx}
\usepackage{amssymb}
\usepackage{times}
\usepackage{anysize}
\usepackage{cite}
\usepackage{lineno}

\marginsize{2cm}{2cm}{1cm}{1cm}
\newtheorem{theorem}{Theorem}

\newtheorem{proposition}[theorem]{Proposition}
\newtheorem{remark}[theorem]{Remark}

\newenvironment{proof}[1][Proof]{\noindent\textbf{#1.} }{\ \rule{0.5em}{0.5em}}
\input{tcilatex}
\begin{document}

\title{Entanglement of Classical and Quantum Short-Range Dynamics in
Mean-Field Systems}
\author{J.-B. Bru\thanks{%
Departamento de Matem\'{a}ticas, Facultad de Ciencia y Tecnolog\'{\i}a,
Universidad del Pa\'{\i}s Vasco, Apartado 644, 48080 Bilbao \& BCAM - Basque
Center for Applied Mathematics, Mazarredo, 14. 48009 Bilbao \& IKERBASQUE,
Basque Foundation for Science, 48011, Bilbao. Email: jb.bru@ikerbasque.org}
\and W. de Siqueira Pedra\thanks{%
Departamento de F\'{\i}sica Matem\'{a}tica, Instituto de F\'{\i}sica,
Universidade de S\~{a}o Paulo, Rua do Mat\~{a}o 1371, CEP 05508-090 S\~{a}o
Paulo, SP Brasil. Email: wpedra@if.usp.br}}
\maketitle

\begin{abstract}
The relationship between classical and quantum mechanics is usually
understood via the limit $\hbar \rightarrow 0$. This is the underlying idea
behind the quantization of classical objects. The apparent incompatibility
of general relativity with quantum mechanics and quantum field theory has
challenged for many decades this basic idea. We recently showed \ \cite%
{Bru-pedra-MF-I, BruPedra-MFII,BruPedra-MFIII} the emergence of classical
dynamics for very general quantum lattice systems with mean-field
interactions, without (complete) supression of its quantum features, in the
infinite volume limit. This leads to a theoretical framework in which the
classical and quantum worlds are entangled. Such an entanglement is
noteworthy and is a consequence of the highly non-local character of
mean-field interactions. Therefore, this phenomenon should not be restricted
to systems with mean-field interactions only, but should also appear in
presence of interactions that are sufficiently long-range, yielding
effective, classical background fields, in the spirit of the Higgs mechanism
of quantum field theory. In order to present the result in a less abstract
way than in its original version, here we apply it to a concrete, physically
relevant, example and discuss, by this means, various important aspects of
our general approach. The model we consider is not exactly solvable and the
particular results obtained are new.\bigskip

\end{abstract}

\section{Introduction}

The limit $\hbar \rightarrow 0$ of Planck's constant refers in mathematics
to the semi-classical analysis, a well-developed and matured research field 
\cite{MR2952218,MR1735654,MR3362292,MR708966,MR1221356,MR3357117}. In
physics, quantum systems are, in many cases, related to classical
counterparts with $\hbar $ appearing as a small deformation parameter, like,
for instance, in Weyl's quantization. See, e.g., \cite[Chapter 13]{quantum
theory}. This is the common understanding\footnote{%
At least in many textbooks on quantum mechanics. See for instance \cite[%
Section 12.4.2, end of the 4th paragraph of page 178]{quantum theory}. In
fact, the nowadays usual correspondence principle (which is, by the way, not
precisely the original principle that Bohr had in mind \cite[Section 4.2]%
{Bokulich}) says that the classical world can appear for large quantum
numbers via a statistical interpretation of quantum mechanics. Nonetheless,
this does not necessarily mean that one has to perform the limit $\hbar
\rightarrow 0$, as Bohr himself stressed. Quoting \cite[p. 313]{Miller}:
\textquotedblleft Edward M. Purcell informed me that Niels Bohr made a
similar comment during a visit to the Physics Department at Harvard
University in 1961. The place was Purcell's office where Purcell and others
had taken Bohr for a few minutes of rest. They were in the midst of a
general discussion when Bohr commented: \textit{People say that classical
mechanics is the limit of quantum mechanics when} $h$ \textit{goes to zero.}%
\ Then, Purcell recalled, Bohr shook his finger and walked to the blackboard
on which he wrote $e^{2}/hc$. As he made three strokes under $h$, Bohr
turned around and said, \textit{you see }$h$ \textit{is in the denominator.}%
\textquotedblright\ A picture of the blackboard can be found in \cite[p. 313]%
{Miller}. See also \cite{Bokulich,Landsman07} and references therein for an
exhaustive discussions on relations between classical and quantum mechanics.}
of the relationship between quantum and classical mechanics, which is
formally seen as a limit case\footnote{%
This limit case $\hbar \rightarrow 0$ corresponds in fact to the so-called
semiclassical mechanics, referring to \textquotedblleft putting quantum
flesh on classical bones\textquotedblright\ \cite[Section 5.1]{Bokulich}.}
of quantum mechanics, even if there exist physical features (such as the
spin of quantum particles) which do not have a clear classical counterpart.
This is reminiscent of the widespread oversight\footnote{%
See for instance \cite[3.2.3 a.]{quantum theory}.} that Planck's
revolutionary ideas to explain thermal radiation in 1900 was not only the
celebrated Planck's constant $\hbar $ (discontinuity of energy), but also
the introduction of an unusual\footnote{%
in regards to Boltzmann's studies, which meanwhile have strongly influenced
Planck's work. In modern terms Planck used the celebrated Bose--Einstein
statistics. In this context, Bose-Einstein condensation, superfluidity and
superconductivity, which may be associated with classical equations, are
consequence of the non-classical statistics of corresponding quantum
particles (bosons or fermions), which is of course extensively verified in
experiments (e.g., in rotating ultracold dilute Bose gases).} statistics
(without any conceptual foundation, in a \textit{ad hoc} way).

Nevertheless, classical mechanics does not only appear in the limit $\hbar
\rightarrow 0$, but also in quantum systems with \emph{mean-field }%
interactions. Theoretical physicists are of course aware of this fact. See,
e.g., \cite{extra-ref0000} where the mean-field (classical) theory
corresponds to the leading term of a \textquotedblleft large $N$%
\textquotedblright\ expansion while the quantum part of the theory (quantum
fluctuations) is related to the next-to-leading order term. This can be
traced back, at least, down to Bogoliubov's microscopic theory of
superfluidity of helium 4 \cite{BruZagrebnov8}.

In 1947, Bogoliubov proposes an ansatz, widely known as the Bogoliubov
approximation, which corresponds to replace, in many-boson Hamiltonians, the
annihilation and creation operators of zero-momentum particles with complex
numbers to be determined \emph{self-consistently}. See \cite[Section 1.1]%
{BruZagrebnov8} for more details. However, even nowadays, the mathematical
validity of this approximation with respect to the primordial dynamics of
(stable) many-boson Hamiltonians with usual two-body interactions is an open
problem\footnote{%
The Bogoliubov approximation should lead to classical equations for the time
evolution of a (classical) field $\{c(t)\}_{t\in \mathbb{R}}\subseteq 
\mathbb{C}$, while the remaining quantum dynamics is supposed to be
generated by a family $\{H(c(t))\}_{t\in \mathbb{R}}$ of time-dependent
Hamiltonians. So, at least heuristically, one obtains two coupled dynamics:
one should be classical, the other one, quantum. It has been recently proven 
\cite{Ammari2018} that the Gross-Pitaevskii and Hartree hierarchies are
equivalent to Liouville's equations for infinite-dimensional functional
spaces. Nevertheless, for these particular systems, the mean-field limits
can be rewritten as semi-classical limits and, as a consequence, no quantum
part appears in the (macroscopic) mean-field dynamics of the associated Bose
gases. See again \cite{Ammari2018} and references therein.}$^{,}$\footnote{%
The validity of the Bogoliubov approximation at \emph{equilibrium} was first
rigorously justified in 1968 by Ginibre \cite{Ginibre} on the level of the
grand-canonical pressure in the thermodynamic limit. See also \cite%
{LiebSeiringerYngvason3}. On the level of states, this question is an old
open problem in mathematical physics, see \cite[p. 28]{Ginibre}.}.

In 1957, the Bogoliubov approximation as well as Bogoliubov's idea of
quasi-particle or \textquotedblleft pairing\textquotedblright\ was adapted
within the framework of electron systems (fermions) in the celebrated
Bardeen-Cooper-Schrieffer (BCS) theory, which explains conventional (type I)
superconductivity. In this context, more precisely for the (exactly
solvable) strong-coupling BCS Hamiltonian, in 1967 Thirring and Wehrl
contributed a first study \cite{T1,T2} on the validity of the Bogoliubov
approximation (understood here as a Hamiltonian) as the effective generator
of this mean-field dynamics in the thermodynamic limit.

In 1973, Hepp and Lieb \cite{Hepp-Lieb73} made explicit, for the first time,
the existence of Poisson brackets in some (commutative) algebra of
functions, related to a \emph{classical} effective dynamics. Hepp and Lieb's
physical motivation was to understand the properties of a laser coupled to a
reservoir and, roughly speaking, in this context, they studied a
permutation-invariant quantum-spin system with mean-field interactions.

The paper \cite{Hepp-Lieb73} is seminal and this research line was further
developed by many other authors, at least until the nineties. See, e.g., 
\cite%
{Bona75,Sewell83,Rieckers84,Morchio87,Bona87,Duffner-Rieckers88,Bona88,Bona89,Bona90,Unnerstall90,Unnerstall90b,Unnerstall90-open,Bona91,Duffield1991,BagarelloMorchio92,Duffield-Werner1,Duffield-Werner2,Duffield-Werner3,Duffield-Werner4,Duffield-Werner5}%
. We focus here on B\'{o}na's approach, referring to his impressive series
of papers, starting in 1975 with \cite{Bona75}. In the middle of the
eighties, his results \cite{Bona83,Bona86} lead him to consider a non-linear
generalization of quantum mechanics. Based on his decisive progresses \cite%
{Bona87,Bona88,Bona89,Bona90} on permutation-invariant quantum-spin systems
with mean-field interactions, B\'{o}na presents a full-fledged abstract
theory\footnote{%
The construction given in the recent paper \cite{Kryukov} for a Hamiltonian
flow associated with Schr\"{o}dinger's dynamics of one quantum particle
corresponds to a particular case of B\'{o}na's theory. However, the author
of \cite{Kryukov} does not seem to be aware of B\'{o}na's works.} in 1991 
\cite{Bona91} and later in a mature textbook published in 2000 (revised in
2012) \cite{Bono2000}, named by him \textquotedblleft extended quantum
mechanics\textquotedblright . See also his new book \cite{Bona2020} on the
subject, published in 2020.

Following \cite[Section 1.1-a]{Bono2000}, B\'{o}na's original motivation was
to \textquotedblleft understand connections between quantum and classical
mechanics more satisfactorily than via the limit $\hbar \rightarrow 0$%
.\textquotedblright\ His major conceptual contribution is to highlight the
emergence of classical mechanics without necessarily the disappearance of
the quantum world, offering a general formal mathematical framework to
understand physical phenomena with macroscopic quantum coherence.

B\'{o}na's view point is different from recent approaches of theoretical
physics like \cite%
{extra-ref1988,extra-ref,extra-ref2,extra-ref3,extra-ref00,extra-ref1} (see
also references therein), which propose a general formalism to get a
consistent description of interactions between classical and quantum
systems, having in mind chemical reactions, decoherence or the quantum
measurement theory. The approaches \cite%
{extra-ref1988,extra-ref,extra-ref2,extra-ref3,extra-ref00,extra-ref1} (see
also references therein) refer to quantum-classical hybrid theories for
which the classical space exists by definition, in a \emph{ad hoc }way,
because of measuring instruments for instance. In fact, many important
models of quantum mechanics already represent systems of quantum particles
in interaction with classical fields. For example, a quantum particle
interacting with an external electromagnetic field is commonly studied via
the magnetic Laplacian. In other words, these models implicitly combine
quantum and classical mechanics. This simplification is physically justified
by the huge numbers of photons giving origin to (effective classical)
macroscopic fields, in the spirit of the correspondence principle. It can
also mathematically be justified, like for instance in the very recent paper 
\cite{3}\footnote{\cite{3} gives a mathematical justification of such a
procedure for three important quantum models: the Nelson, Pauli-Fierz and
Polaron models. Mathematically, it refers to a semi-classical analysis of
the bosonic degrees of freedom, leading to a new (equivalent) quantum model
interacting with a classical field. Thanks to \cite{1,2}, the authors are
able to handle the semi-classical limit for very general (possibly
entangled, i.e., not factorized) states.}. By contrast, in B\'{o}na's view
point, the classical world emerges \emph{intrinsically} from macroscopic
quantum systems, like in \cite{extra-ref0bis}. This is also similar to \cite%
{extra-ref0}, which is however a much more elementary example\footnote{%
It corresponds to a quantum systems with two species of particles in an
extreme mass ratio limit: one species becomes, in this limit, infinitely
more massive than the other one. In this limit, the massive species, like
nuclei, becomes classical while the other one, like electrons, stays quantum.%
} referring to the Ehrenfest dynamics.

In \cite{Bru-pedra-MF-I} we revisit B\'{o}na's conceptual lines, but propose
a new method to mathematically implement them, with a much broader domain of
applicability than his original version \cite{Bono2000} (see also \cite%
{landsmann07,Odzijewicz,Bona2020} and references therein). In fact, similar
to B\'{o}na, who constructs a general abstract theory \cite{Bono2000} based
on previous progress \cite{Bona87,Bona88,Bona89,Bona90} on
permutation-invariant quantum-spin systems with mean-field interactions, we
also base our abstract theory \cite{Bru-pedra-MF-I} on our own (completely
new) results \cite{BruPedra-MFII,BruPedra-MFIII,BruPedra-MFVI} on the
dynamical properties of (possibly non-permutation-invariant)\ quantum
lattice systems with mean-field interactions. Our approach gives, in the
infinite volume limit, an explicit representation of the full dynamics of
such systems as entangled classical and quantum short-range dynamics. In
particular, in contrast to B\'{o}na's one, we highlight the relation between
the phase space of the corresponding classical dynamics and the state space
of the non-commutative algebra where the quantum short-range dynamics
meanwhile runs, making meanwhile explicit the central role played by \emph{%
self-consistency}.

The general theory can be found in \cite{Bru-pedra-MF-I}, which is a rather
long mathematical paper (72 pages). The aim of the current paper is thus to
illustrate, in a simple manner, the entanglement of classical and quantum
short-range dynamics, as well as important aspects our the general approach.
This is done via the so-called strong-coupling BCS-Hubbard model, which
serves here as a paradigm. From a technical viewpoint, the dynamical
properties of this model are easy to study, albeit non-trivial, the model
being not exactly solvable. From the physical point of view, this model is
also interesting because it highlights the possible thermodynamic impact of
the (screened) Coulomb repulsion on ($s$-wave) superconductivity, in the
strong-coupling approximation. Its behavior at thermodynamical equilibrium
is already rigorously known \cite{BruPedra1}, but \emph{not} its infinite
volume dynamics. In fact, note that \cite{Bru-pedra-proceeding} is merely a
concise introduction to this problem and the particular results presented
here are new. The precise definition of the model and all its dynamical
properties are explained in Section \ref{Strong-Coupling BCS-Hubbard}. The
entanglement of classical and quantum (short-range) dynamics in this
prototypical model is then made explicit in Section \ref{Section
Entanglement}.

For convenience of the reader interested in the mathematical results \cite%
{BruPedra-MFII,BruPedra-MFIII} on the macroscopic dynamics of\ fermion and
quantum-spin systems with mean-field interactions, we provide an appendix,
since \cite{BruPedra-MFII,BruPedra-MFIII} are altogether about 126 pages
long. Appendix \ref{Mathematical Foundations} explains \cite%
{BruPedra-MFII,BruPedra-MFIII} in concise, albeit mathematically precise,
terms. In Appendix \ref{sect Permutation-Invariant Long-Range Models}, note
that we formulate the results in the special context of
permutation-invariant models, making the link with the strong-coupling
BCS-Hubbard model and previous results on permutation-invariant quantum-spin
systems. Appendix \ref{sect Permutation-Invariant Long-Range Models}
contains \emph{new} material that cannot be found in our previous papers 
\cite%
{BruPedra-MFII,BruPedra-MFIII,Bru-pedra-proceeding,BruPedra-MFVI,BruPedra1}
on the subject.

\begin{remark}
In all the paper, we focus on lattice-fermion systems, but all the results
and discussions can be translated to quantum-spin systems via obvious
modifications.
\end{remark}

\section{The Strong-Coupling BCS-Hubbard Model\label{Strong-Coupling
BCS-Hubbard}}

\subsection{Presentation of the Model\label{Presentation of the Model}}

The dynamics of the (reduced) BCS\ Hamiltonian can be \emph{explicitly}
computed by means of \cite{BruPedra-MFII,BruPedra-MFIII}, but we prefer to
consider here a BCS-type model including the Hubbard interaction. In fact,
it is a much richer new example while the BCS\ Hamiltonian was already been
extensively studied in the literature, via various approaches. Observe, in
particular, that the (usual, reduced) strong-coupling BCS model is exactly
solvable, whereas its extension considered here is not. We call this new
model the strong-coupling BCS-Hubbard Hamiltonian. Its equilibrium states
were rigorously studied in \cite{BruPedra1}, in order to understand the
possible thermodynamic impact of the Coulomb repulsion on ($s$-wave)
superconductivity. An interesting outcome of \cite{BruPedra1} is a
mathematically rigorous proof of the existence of a superconductor-Mott
insulator phase transition for the strong-coupling BCS-Hubbard Hamiltonian,
like in cuprates, which must be doped in order to avoid the insulating
(Mott) phase and become superconductors.

The results of \cite{BruPedra1} refer to an \emph{exact} study of the phase
diagram of the strong-coupling BCS-Hubbard model, whose Hamiltonian is
defined in any cubic box $\Lambda _{L}:=\{\mathbb{Z}\cap \left[ -L,L\right]
\}^{d}$ ($d\in \mathbb{N}$) of volume $|\Lambda _{L}|$, $L\in \mathbb{N}_{0}$%
, by%
\begin{equation}
\mathrm{H}_{L}:=\sum\limits_{x\in \Lambda _{L}}h_{x}-\frac{\gamma }{%
\left\vert \Lambda _{L}\right\vert }\sum_{x,y\in \Lambda _{L}}a_{x,\uparrow
}^{\ast }a_{x,\downarrow }^{\ast }a_{y,\downarrow }a_{y,\uparrow }
\label{strong coupling ham}
\end{equation}%
for real parameters $\mu ,h\in \mathbb{R}$ and $\lambda ,\gamma \geq 0$,
where, for all $x\in \mathbb{Z}^{d}$,%
\begin{eqnarray}
&&h_{x}%
:=%
2\lambda n_{x,\uparrow }n_{x,\downarrow }-\mu \left( n_{x,\uparrow
}+n_{x,\downarrow }\right)  \label{strong coupling hamb} \\
&&\qquad \qquad \qquad \qquad \qquad -h\left( n_{x,\uparrow
}-n_{x,\downarrow }\right) .  \notag
\end{eqnarray}%
Recall that the operator $a_{x,\mathrm{s}}^{\ast }$ ($a_{x,\mathrm{s}}$)
creates (annihilates) a fermion with spin $\mathrm{s}\in \{\uparrow
,\downarrow \}$ at lattice position $x\in \mathbb{Z}^{d}$, $d=1,2,3,...$,
whereas $n_{x,\mathrm{s}}:=a_{x,\mathrm{s}}^{\ast }a_{x,\mathrm{s}}$ is the
particle number operator at position $x$ and spin $\mathrm{s}$. They are
linear operators acting on the fermion Fock space $\mathcal{F}_{\Lambda
_{L}} $, where%
\begin{equation}
\mathcal{F}_{\Lambda }:=\bigwedge \mathbb{C}^{\Lambda \times \{\uparrow
,\downarrow \}}\equiv \mathbb{C}^{2^{\Lambda \times \{\uparrow ,\downarrow
\}}}  \label{fock}
\end{equation}%
for any $\Lambda \subseteq \mathbb{Z}^{d}$ and $d\in \mathbb{N}$.

The first term of the right-hand side of (\ref{strong coupling hamb})
represents the (screened) Coulomb repulsion as in the celebrated Hubbard
model. The second term corresponds to the strong-coupling limit of the
kinetic energy, also called \textquotedblleft atomic
limit\textquotedblright\ in the Hubbard model community, the real parameter $%
\mu $ being the so-called chemical potential. The third term is the
interaction between spins and the external magnetic field $h$.

The last term in (\ref{strong coupling ham}) is the (homogeneous) BCS
interaction written in the position space (see, e.g., \cite[Eq. (1.3)]%
{BruPedra1}). The \emph{long-range} character of this interaction is
apparent for it is an infinite-range hopping term (for fermion pairs). In
fact, it is a mean-field interaction, since%
\begin{eqnarray*}
&&\frac{1}{\left\vert \Lambda _{L}\right\vert }\sum_{x,y\in \Lambda
_{L}}a_{x,\uparrow }^{\ast }a_{x,\downarrow }^{\ast }a_{y,\downarrow
}a_{y,\uparrow } \\
&=&\sum_{y\in \Lambda _{L}}\left( \frac{1}{\left\vert \Lambda
_{L}\right\vert }\sum_{x\in \Lambda _{L}}a_{x,\uparrow }^{\ast
}a_{x,\downarrow }^{\ast }\right) a_{y,\downarrow }a_{y,\uparrow }\ .
\end{eqnarray*}%
This is a simple example of the far more general case studied in \cite%
{BruPedra-MFII,BruPedra-MFIII}. It is however a non-trivial and a very
interesting mean-field model since, even when $\mu =h=\lambda =0$, the
Hamiltonian $\mathrm{H}_{L}$ qualitatively displays most of basic properties
of real conventional type I superconductors. See, e.g. \cite[Chapter VII,
Section 4]{Thou}. Note that the precise mediators leading to the effective
BCS interaction are not relevant here, i.e., they could be phonons, as in
conventional type I superconductors, or anything else.

\subsection{Approximating Hamiltonians}

The thermodynamic impact of the Coulomb repulsion on $s$-wave
superconductors is analyzed in \cite{BruPedra1}, via a rigorous study of
equilibrium and ground states of the strong-coupling BCS-Hubbard
Hamiltonian: At any $L_{0}\in \mathbb{N}_{0}$ and inverse temperature $\beta
>0$, for any linear operator $A$ acting on the fermion Fock space $\mathcal{F%
}_{\Lambda _{L_{0}}}$ (\ref{fock}), we prove that 
\begin{equation}
\lim_{L\rightarrow \infty }\omega ^{(L)}\left( A\right) =\omega \left(
A\right) ,  \label{weak*}
\end{equation}%
where, for $L\in \mathbb{N}_{0}$, 
\begin{equation}
\omega ^{(L)}\left( \cdot \right) :=\mathrm{Trace}_{\mathcal{F}_{\Lambda
_{L}}}\left( \left( \cdot \right) \frac{\mathrm{e}^{-\beta \mathrm{H}_{L}}}{%
\mathrm{Trace}_{\mathcal{F}_{\Lambda _{L}}}\left( \mathrm{e}^{-\beta \mathrm{%
H}_{L}}\right) }\right)  \label{gibbs1}
\end{equation}%
is the Gibbs states associated with $\mathrm{H}_{L}$, while $\omega $ is an 
\emph{explicitly given }(infinite volume) equilibrium state, defined as
being a (global, space-homogeneous) minimizer of the free energy density
(i.e., free energy per unit volume). See \cite[Section 6.2]{BruPedra1} for
more details.

An important point in such an analysis is the study of an associate
variational problem over complex numbers: By the so-called approximating
Hamiltonian method \cite%
{approx-hamil-method0,approx-hamil-method,approx-hamil-method2}, one defines
an approximation of the Hamiltonian, which is, in the case of the
strong-coupling BCS-Hubbard Hamiltonian, the $c$-dependent Hamiltonian%
\begin{equation}
\mathrm{H}_{L}\left( c\right) :=\sum\limits_{x\in \Lambda _{L}}\left\{
h_{x}-\gamma \left( ca_{x,\uparrow }^{\ast }a_{x,\downarrow }^{\ast }+\bar{c}%
a_{x,\downarrow }a_{x,\uparrow }\right) \right\}
\label{Hamiltonian BCS-Hubbard approx}
\end{equation}%
where $c\in \mathbb{C}$. The main advantage of using this $c$-dependent
Hamiltonian, in comparison with $\mathrm{H}_{L}$, is the fact that it is a
sum of shifts of the same on-site operator. For an appropriate choice of
(order) parameter $c\in \mathbb{C}$, it leads to the exact pressure of the
strong-coupling BCS-Hubbard model, in the limit $L\rightarrow \infty $: At
inverse temperature $\beta >0$, 
\begin{equation}
\lim_{L\rightarrow \infty }p\left[ \mathrm{H}_{L}\right] =\underset{c\in 
\mathbb{C}}{\sup }\left\{ -\gamma |c|^{2}+\lim_{L\rightarrow \infty }p\left[ 
\mathrm{H}_{L}\left( c\right) \right] \right\}  \label{var pb}
\end{equation}%
with $p\left[ \mathrm{H}\right] $ being the pressure%
\begin{equation}
p\left[ \mathrm{H}\right] :=\frac{1}{\beta \left\vert \Lambda
_{L}\right\vert }\ln \mathrm{Trace}_{\mathcal{F}_{\Lambda _{L}}}\left( 
\mathrm{e}^{-\beta \mathrm{H}}\right) ,\qquad \beta >0,  \notag
\end{equation}%
associated with any Hamiltonian $\mathrm{H}$ acting on the fermion Fock
space $\mathcal{F}_{\Lambda _{L}}$. In fact, the (exact) Gibbs state $\omega
^{(L)}$ converges\footnote{%
In the sense of (\ref{weak*}), or, in the mathematical jargon, in the weak$%
^{\ast }$ topology.} to a convex combination of the thermodynamic limit $%
L\rightarrow \infty $ of the (approximating) Gibbs state $\omega ^{(L,%
\mathfrak{d})}$ defined by 
\begin{equation}
\omega ^{(L,\mathfrak{d})}\left( \cdot \right) :=\mathrm{Trace}_{\mathcal{F}%
_{\Lambda _{L}}}\left( \left( \cdot \right) \frac{\mathrm{e}^{-\beta \mathrm{%
H}_{L}\left( \mathfrak{d}\right) }}{\mathrm{Trace}_{\mathcal{F}_{\Lambda
_{L}}}\left( \mathrm{e}^{-\beta \mathrm{H}_{L}\left( \mathfrak{d}\right)
}\right) }\right) ,  \label{gibbs2}
\end{equation}%
the complex number $\mathfrak{d}\in \mathbb{C}$ being a solution to the
variational problem (\ref{var pb}).

Since $\gamma \geq 0$, this can heuristically be seen from the inequality 
\begin{eqnarray*}
&&\gamma \left\vert \Lambda _{L}\right\vert \left\vert c\right\vert ^{2}+%
\mathrm{H}_{L}\left( c\right) -\mathrm{H}_{L} \\
&=&\gamma \left( \mathfrak{c}_{0}^{\ast }-\sqrt{\left\vert \Lambda
_{L}\right\vert }\bar{c}\right) \left( \mathfrak{c}_{0}-\sqrt{\left\vert
\Lambda _{L}\right\vert }c\right) \geq 0,
\end{eqnarray*}%
where 
\begin{equation}
\mathfrak{c}_{0}:=\frac{1}{\sqrt{\left\vert \Lambda _{L}\right\vert }}%
\sum_{x\in \Lambda _{L}}a_{x,\downarrow }a_{x,\uparrow }
\label{dynamics approx00}
\end{equation}%
($\mathfrak{c}_{0}^{\ast }$) annihilates (creates) one Cooper pair within
the condensate, i.e., in the zero-mode for fermion pairs. This suggests the
(rigorously proven) fact \cite[Theorem 3.1]{BruPedra1} that 
\begin{equation}
\left\vert \mathfrak{d}\right\vert ^{2}=\lim_{L\rightarrow \infty }\frac{%
\omega ^{(L)}\left( \mathfrak{c}_{0}^{\ast }\mathfrak{c}_{0}\right) }{%
\left\vert \Lambda _{L}\right\vert }  \label{dynamics approx0}
\end{equation}%
for any\footnote{%
This implies that any solution $\mathfrak{d}$ to the variational problem (%
\ref{var pb}) must have the same absolute value.} $\mathfrak{d}\in \mathbb{C}
$ solution to the variational problem (\ref{var pb}). The parameter $%
\left\vert \mathfrak{d}\right\vert ^{2}$ is the condensate density of Cooper
pairs and so, $\mathfrak{d}\neq 0$ corresponds to the existence of a
superconducting phase, which is shown to exist for sufficiently large $%
\gamma \geq 0$. See also \cite[Figs. 1,2,3]{BruPedra1}.

\subsection{Dynamical Problem}

As is usual, an Hamiltonian like the strong-coupling BCS-Hubbard Hamiltonian
drives a dynamics in the Heisenberg picture of quantum mechanics: The
corresponding time-evolution is, for $L\in \mathbb{N}_{0}$, a continuous
group $(\tau _{t}^{(L)})_{t\in {\mathbb{R}}}$ of automorphisms of the
algebra $\mathcal{B}(\mathcal{F}_{\Lambda _{L}})$ of linear operators acting
on the fermion Fock space $\mathcal{F}_{\Lambda _{L}}$ (see (\ref{fock})),
defined by 
\begin{equation}
\tau _{t}^{(L)}(A):=\mathrm{e}^{it\mathrm{H}_{L}}A\mathrm{e}^{-it\mathrm{H}%
_{L}}  \label{dynamics full}
\end{equation}%
for any $A\in \mathcal{B}(\mathcal{F}_{\Lambda _{L}})$ and $t\in {\mathbb{R}}
$. The generator\footnote{%
That is, $\tau _{t}^{(L)}(A)=\exp \left( it\delta _{L}\right) $.} of this
time evolution is the linear operator $\delta _{L}$ defined on $\mathcal{B}(%
\mathcal{F}_{\Lambda _{L}})$ by%
\begin{equation*}
\delta _{L}\left( A\right) :=i[\mathrm{H}_{L},A]:=i\left( \mathrm{H}_{L}A-A%
\mathrm{H}_{L}\right) .
\end{equation*}

If $\gamma =0$ then, for any time $t\in {\mathbb{R}}$ and linear operator $A$
acting on the fermion Fock space $\mathcal{F}_{\Lambda _{L_{0}}}$ (\ref{fock}%
), $L_{0}\in \mathbb{N}_{0}$,%
\begin{equation}
\begin{array}{l}
\lim_{L\rightarrow \infty }\tau _{t}^{(L)}\left( A\right) =\tau
_{t}^{(L_{0})}\left( A\right) , \\ 
\lim_{L\rightarrow \infty }\delta _{L}\left( A\right) =i[\mathrm{H}%
_{L_{0}},A],%
\end{array}
\label{limit}
\end{equation}%
because $\mathrm{H}_{L}|_{\gamma =0}$ is the sum of on-site terms. In
particular, (\ref{limit}) uniquely defines an infinite volume dynamics in
this case. Nonetheless, as soon as $\gamma >0$, the thermodynamic limit (\ref%
{limit}) of the mean-field dynamics does \emph{not} exists in general (even
along subsequences).

One can try to approximate $\tau _{t}^{(L)}$ by $\tau _{t}^{(L,c)}$, where 
\begin{equation}
\tau _{t}^{(L,c)}(A):=\mathrm{e}^{it\mathrm{H}_{L}\left( c\right) }A\mathrm{e%
}^{-it\mathrm{H}_{L}\left( c\right) }  \label{dynamics approx}
\end{equation}%
for any $L\in \mathbb{N}_{0}$, $A\in \mathcal{B}(\mathcal{F}_{\Lambda _{L}})$
and some complex number $c\in \mathbb{C}$. In this case, the linear operator 
\begin{equation}
\delta _{L,c}\left( \cdot \right) :=i[\mathrm{H}_{L}\left( c\right) ,\cdot ]
\label{generator approx}
\end{equation}%
on $\mathcal{B}(\mathcal{F}_{\Lambda _{L}})$ is the generator of the
dynamics $(\tau _{t}^{(L,c)})_{t\in {\mathbb{R}}}$. In this case, since
local Hamiltonians (\ref{Hamiltonian BCS-Hubbard approx}) are sums of
on-site terms, for any $c\in {\mathbb{C}}$, $t\in {\mathbb{R}}$, $L_{0}\in 
\mathbb{N}_{0}$ and $A\in \mathcal{F}_{\Lambda _{L_{0}}}$,%
\begin{equation}
\begin{array}{l}
\lim_{L\rightarrow \infty }\tau _{t}^{(L,c)}\left( A\right) =\tau
_{t}^{(L_{0})}\left( A\right) , \\ 
\lim_{L\rightarrow \infty }\delta _{L,c}\left( A\right) =i[\mathrm{H}%
_{L_{0}}(c),A],%
\end{array}
\label{limit2}
\end{equation}%
like in the case $\gamma =0$ with (\ref{limit}). In other words, there is an
infinite volume dynamics for such approximating interactions.

A natural choice for $c\in \mathbb{C}$ would be a solution to the
variational problem (\ref{var pb}), but what about if the solution is not
unique?\ Observe, moreover, that the variational problem (\ref{var pb})
depends on the temperature whereas the time evolution (\ref{dynamics full})
does not!

The validity of the approximation with respect to the primordial dynamics
was an open question that Thirring and Wehrl \cite{T1,T2} solve in 1967 for
the special case $\mathrm{H}_{L}|_{\mu =\lambda =h=0}$, which is an exactly
solvable permutation-invariant model for any $\gamma \in \mathbb{R}$. An
attempt to generalize Thirring and Wehrl's results to a general class of
fermionic models, including the BCS theory, has been done in 1978 \cite%
{Hemmen78}, but at the cost of technical assumptions that are difficult to
verify in practice.\ This research direction has been strongly developed by
many authors until 1992, see \cite%
{Bona75,Sewell83,Rieckers84,Morchio87,Bona87,Duffner-Rieckers88,Bona88,Bona89,Bona90,Unnerstall90,Unnerstall90b,Unnerstall90-open,Bona91,Duffield1991,BagarelloMorchio92,Duffield-Werner1,Duffield-Werner2,Duffield-Werner3,Duffield-Werner4,Duffield-Werner5}%
. All these papers study dynamical properties of \emph{permutation-invariant}
quantum-spin systems with mean-field interactions. Our results \cite%
{BruPedra-MFII,BruPedra-MFIII}, summarized in Appendix \ref{Mathematical
Foundations}, represent a significant generalization of such previous
results to possibly non-permutation-invariant lattice-fermion or
quantum-spin systems. In order to illustrate how our results \cite%
{BruPedra-MFII,BruPedra-MFIII} are used to control the infinite volume
dynamics of mean-field Hamiltonians, we now come back to our pedagogical
example, that is, the strong-coupling BCS-Hubbard model.

\subsection{Dynamical Self-Consistency}

Instead of considering the Heisenberg picture, let us consider the Schr\"{o}%
dinger picture of quantum mechanics. In this case, recall that, at fixed $%
L\in \mathbb{N}_{0}$, a (finite volume) state $\rho ^{(L)}$ is a positive
and normalized functional acting on the algebra $\mathcal{B}(\mathcal{F}%
_{\Lambda _{L}})$ of linear operators on the fermion Fock space $\mathcal{F}%
_{\Lambda _{L}}$. By finite dimensionality of $\mathcal{F}_{\Lambda _{L}}$, 
\begin{equation*}
\rho ^{(L)}\left( \cdot \right) :=\mathrm{Trace}_{\mathcal{F}_{\Lambda
_{L}}}\left( \left( \cdot \right) \mathrm{d}^{(L)}\right) ,
\end{equation*}%
for a uniquely defined positive operator $\mathrm{d}^{(L)}\in \mathcal{B}(%
\mathcal{F}_{\Lambda _{L}})$ satisfying $\mathrm{Trace}_{\mathcal{F}%
_{\Lambda _{L}}}(\mathrm{d}^{(L)})=1$ and named the density matrix of $\rho
^{(L)}$. Compare with (\ref{gibbs1}) and (\ref{gibbs2}). See also Appendix %
\ref{Finite-Volume State Spaces}. At $L\in \mathbb{N}_{0}$, the expectation
of any $A\in \mathcal{B}(\mathcal{F}_{\Lambda _{L}})$ at time $t\in {\mathbb{%
R}}$ is, as usual, equal to%
\begin{equation}
\rho _{t}^{(L)}\left( A\right) :=\mathrm{Trace}_{\mathcal{F}_{\Lambda
_{L}}}\left( \mathrm{e}^{it\mathrm{H}_{L}}A\mathrm{e}^{-it\mathrm{H}_{L}}%
\mathrm{d}^{(L)}\right) .  \label{rhorho}
\end{equation}%
I.e., the time evolution of any finite volume state is 
\begin{equation}
\rho _{t}^{(L)}:=\rho ^{(L)}\circ \tau _{t}^{(L)}\ ,\qquad t\in {\mathbb{R}}%
\ ,  \label{rho}
\end{equation}%
which corresponds to a time-dependent density matrix equal to $\mathrm{d}%
_{t}^{(L)}=\tau _{-t}^{(L)}(\mathrm{d}^{(L)})$. Compare with (\ref%
{long-range dyn0}).

The thermodynamic limit of (\ref{rhorho}) for periodic states can be
explicitly computed, as explained in Appendix \ref{Quantum Part of
Long-Range Dynamics}. It refers to a \emph{non}-autonomous state-dependent
dynamics related to \emph{self-consistency}: By (\ref{fock}) with $\Lambda
=\Lambda _{0}=\{0\}$, recall that%
\begin{equation}
\mathcal{F}_{\{0\}}:=\bigwedge \mathbb{C}^{\{0\}\times \{\uparrow
,\downarrow \}}\equiv \mathbb{C}^{4}  \label{Fock2}
\end{equation}%
is the fermion Fock space associated with the lattice site $(0,\ldots ,0)\in 
\mathbb{Z}^{d}$ and so, $\mathcal{B}\left( \mathcal{F}_{\{0\}}\right) $ can
be identified with the algebra $\mathrm{Mat}(4,\mathbb{C})$ of complex $%
4\times 4$ matrices, in some orthonormal basis\footnote{%
For instance, $\left( 1,0,0,0\right) $ is the vacuum; $\left( 0,1,0,0\right) 
$ and $\left( 0,0,1,0\right) $ correspond to one fermion with spin $\uparrow 
$ and $\downarrow $, respectively; $\left( 0,0,0,1\right) $ refers to two
fermions with opposite spins.}. For any continuous family $\omega :=(\omega
_{t})_{t\in \mathbb{R}}$ of states acting on $\mathcal{B}\left( \mathcal{F}%
_{\{0\}}\right) $, we define the (infinite volume) \emph{non}-autonomous
dynamics $(\tau _{t,s}^{(\omega )})_{_{s,t\in \mathbb{R}}}$ by the
Dyson-Phillips series\footnote{%
That is, $\tau _{t,s}^{(\omega )}$ is a time-ordered exponential.} 
\begin{align}
& \tau _{t,s}^{(\omega )}%
:=%
\mathbf{1}+\sum\limits_{k\in {\mathbb{N}}}\int_{s}^{t}\mathrm{d}t_{1}\cdots
\int_{s}^{t_{k-1}}\mathrm{d}t_{k}\text{ }  \label{idiot} \\
& \quad \quad \quad \quad \quad \quad \quad \quad \quad \quad \quad \quad
\delta ^{\omega _{t_{k}}}\circ \cdots \circ \delta ^{\omega _{t_{1}}}  \notag
\end{align}%
where 
\begin{equation}
\delta ^{\rho }:=\lim_{L\rightarrow \infty }\delta _{L,\rho (a_{0,\uparrow
}a_{0,\downarrow })}\left( \cdot \right)  \label{idiot2}
\end{equation}%
is the generator of the infinite volume dynamics associated with the
approximating Hamiltonian $\mathrm{H}_{L}(c)$ for $c=\rho (a_{0,\uparrow
}a_{0,\downarrow })$. See (\ref{Hamiltonian BCS-Hubbard approx}) and (\ref%
{limit2}). Note that the precise definition of the generator $\delta ^{\rho
} $ -- both acting on the CAR algebra $\mathcal{U}$ of the infinite lattice
-- is not necessary here to understand the action of the mappings (\ref%
{idiot})-(\ref{idiot2}) on local elements $A\in \mathcal{B}(\mathcal{F}%
_{\Lambda _{L_{0}}})$, $L_{0}\in \mathbb{N}_{0}$, since in this case 
\begin{equation*}
\delta ^{\rho }\left( A\right) =i[\mathrm{H}_{L_{0}}(\rho (a_{0,\uparrow
}a_{0,\downarrow })),A],
\end{equation*}%
thanks to Equation of (\ref{limit2}). In particular, 
\begin{equation*}
\tau _{t,s}^{(\omega )}\left( \mathcal{B}(\mathcal{F}_{\Lambda
_{L_{0}}})\right) \subseteq \mathcal{B}(\mathcal{F}_{\Lambda
_{L_{0}}}),\qquad L_{0}\in \mathbb{N}_{0}.
\end{equation*}%
Observe that the particular value $\rho (a_{0,\uparrow }a_{0,\downarrow
})\in \mathbb{C}$, which is taken here for the complex parameter $c$, is
reminiscent of (\ref{dynamics approx00})-(\ref{dynamics approx0}).

Now, by (\ref{self-consistency equation}) and (\ref{eq restricted0}), for
any fixed initial (even) state $\rho _{0}$ on $\mathcal{B}\left( \mathcal{F}%
_{\{0\}}\right) $ at $t=0$, there is a unique family $(\mathbf{\varpi }%
(t;\rho _{0}))_{t\in \mathbb{R}}$ of on-site states acting on $\mathcal{B}%
\left( \mathcal{F}_{\{0\}}\right) $ such that 
\begin{equation}
\mathbf{\varpi }(t;\rho _{0})=\rho _{0}\circ \tau _{t,0}^{\mathbf{\varpi }%
(\cdot ;\rho _{0})}\ ,\qquad t\in {\mathbb{R}}\ .  \label{self-consitency}
\end{equation}%
This is a self-consistency equation on a finite-dimensional space, by (\ref%
{Fock2}).

\subsection{Infinite Volume Dynamics for Product States}

For simplicity, at initial time $t=0$, take a finite volume product\footnote{%
The product state $\rho ^{(L)}$ is well-defined by $\rho ^{(L)}(\alpha
_{x_{1}}(A_{1})\cdots \alpha _{x_{n}}(A_{n}))=\rho (A_{1})\cdots \rho
(A_{n}) $ for all $A_{1},\ldots ,A_{n}\in \mathcal{B}\left( \mathcal{F}%
_{\{0\}}\right) $ and all $x_{1},\ldots ,x_{n}\in \Lambda _{L}$ such that $%
x_{i}\not=x_{j}$ for $i\not=j$, where $\alpha _{x_{j}}(A_{j})\in \mathcal{B}%
\left( \mathcal{F}_{\{x_{j}\}}\right) $ is the $x_{j}$-translated copy of $%
A_{j}$ for all $j\in \{1,\ldots ,n\}$. See (\ref{transl}) and (\ref{product
state extremes0})-(\ref{product states}) for more details.} state $\otimes
_{\Lambda _{L}}\rho $ associated with a \emph{fixed} even\footnote{%
Even means that the expectation value of any odd monomials in $\{a_{0,%
\mathrm{s}}^{\ast },a_{0,\mathrm{s}}\}_{\mathrm{s}\in \{\uparrow ,\downarrow
\}}$ with respect to the on-site state $\rho $ is zero. Even states are the
physically relevant ones.} state $\rho $ on $\mathcal{B}\left( \mathcal{F}%
_{\{0\}}\right) $. An example of finite volume product states is given by
the approximating Gibbs states (\ref{gibbs2}). Then, in this case, as
explained in Appendix \ref{sect Permutation-Invariant Long-Range Models},
for any $t\in \mathbb{R}$, the thermodynamic limit 
\begin{equation}
\rho _{t}\left( A\right) :=\lim_{L\rightarrow \infty }\left( \otimes
_{\Lambda _{L}}\rho \right) \circ \tau _{t}^{(L)}\left( A\right)
\label{eq restrictedsimple0}
\end{equation}%
of the expectation of any linear operator $A\in \mathcal{B}(\mathcal{F}%
_{\Lambda _{L_{0}}})$ for $L_{0}\in \mathbb{N}_{0}$ exists and corresponds
to the time-dependent product\footnote{%
For any even state $\tilde{\rho}$ on $\mathcal{B}\left( \mathcal{F}%
_{\{0\}}\right) $, $\otimes _{\mathbb{Z}^{d}}\tilde{\rho}$ is a state acting
on the CAR algebra of the infinite lattice, which includes all $\mathcal{B}(%
\mathcal{F}_{\Lambda _{L}})$, $L\in \mathbb{N}_{0}$. The restriction of $%
\otimes _{\mathbb{Z}^{d}}\tilde{\rho}$ to $\mathcal{B}(\mathcal{F}_{\Lambda
_{L}})$ is of course equal to $\otimes _{\Lambda _{L}}\tilde{\rho}$.} state 
\begin{equation}
\rho _{t}=\left( \otimes _{\mathbb{Z}^{d}}\rho \right) \circ \tau _{t,0}^{%
\mathbf{\varpi }(\cdot ;\rho )}=\otimes _{\mathbb{Z}^{d}}\mathbf{\varpi }%
(t;\rho _{0}),  \label{eq restrictedsimple}
\end{equation}%
where $\mathbf{\varpi }(\cdot ;\rho )$ is defined by (\ref{self-consitency}%
). In other words, for any time $t\in \mathbb{R}$, the limit state is in
this case completely determined by its restriction to the single lattice
site $(0,\ldots ,0)\in \mathbb{Z}^{d}$. Below, we give the explicit time
evolution of the most important physical quantities related to this model,
in this situation:

\begin{proposition}[Infinite volume dynamics]
\label{lemma dynamics}\mbox{ }\newline
\emph{(i) }Electron density:%
\begin{multline*}
\mathrm{d}(\rho )%
:=%
\rho \left( n_{0,\uparrow }+n_{0,\downarrow }\right) =\rho _{t=0}\left(
n_{0,\uparrow }+n_{0,\downarrow }\right) \\
=\rho _{t}\left( n_{0,\uparrow }+n_{0,\downarrow }\right) \in \lbrack 0,2].
\end{multline*}%
\emph{(ii) }Magnetization density: 
\begin{multline*}
\mathrm{m}(\rho )%
:=%
\rho \left( n_{0,\uparrow }-n_{0,\downarrow }\right) =\rho _{t=0}\left(
n_{0,\uparrow }-n_{0,\downarrow }\right) \\
=\rho _{t}\left( n_{0,\uparrow }-n_{0,\downarrow }\right) \in \lbrack -1,1].
\end{multline*}%
\emph{(iii) }Coulomb correlation density: 
\begin{multline*}
\mathrm{w}(\rho )%
:=%
\rho \left( n_{0,\uparrow }n_{0,\downarrow }\right) =\rho _{t=0}\left(
n_{0,\uparrow }n_{0,\downarrow }\right) \\
=\rho _{t}\left( n_{0,\uparrow }n_{0,\downarrow }\right) \in \lbrack 0,1].
\end{multline*}%
\emph{(iv) }Cooper-field and condensate densities:%
\begin{equation*}
\rho _{t}\left( a_{0,\downarrow }a_{0,\uparrow }\right) =\sqrt{\mathrm{%
\kappa }(\rho )}\mathrm{e}^{i\left( t\mathrm{\nu }(\rho )+\theta (\rho
)\right) }
\end{equation*}%
with 
\begin{equation*}
\mathrm{\nu }(\rho ):=2\left( \mu -\lambda \right) +\gamma \left( 1-\mathrm{d%
}(\rho )\right)
\end{equation*}%
and $\mathrm{\kappa }(\rho )\in \lbrack 0,1]$, $\theta (\rho )\in \lbrack
-\pi ,\pi )$ such that, at initial time, $\rho \left( a_{0,\downarrow
}a_{0,\uparrow }\right) =\sqrt{\mathrm{\kappa }(\rho )}\mathrm{e}^{i\theta
(\rho )}$.
\end{proposition}

\begin{proof}
Recall that $[A,B]:=AB-BA$ is the commutator of $A$ and $B$ and, for any $%
\mathrm{s}\in \{\uparrow ,\downarrow \}$, $n_{\mathrm{s}}:=a_{\mathrm{s}%
}^{\ast }a_{\mathrm{s}}$ is the spin-$\mathrm{s}$ particle number operator
on the lattice site $0$ (with \textquotedblleft $0$\textquotedblright\ being
omitted in the notation for simplicity). We now prove Assertions (i)-(iv):
By the canonical anti-commutation relations (CAR), for any $\mathrm{s}\in
\{\uparrow ,\downarrow \}$, 
\begin{equation*}
\left[ dh\left( \rho \right) ,n_{\mathrm{s}}\right] =\gamma \left( \rho
\left( a_{\downarrow }a_{\uparrow }\right) a_{\uparrow }^{\ast
}a_{\downarrow }^{\ast }-\rho \left( a_{\uparrow }^{\ast }a_{\downarrow
}^{\ast }\right) a_{\downarrow }a_{\uparrow }\right)
\end{equation*}%
with $dh\left( \rho \right) $ defined from (\ref{strong coupling hamb}) by 
\begin{equation}
dh\left( \rho \right) :=h_{0}-\gamma \left( a_{\uparrow }^{\ast
}a_{\downarrow }^{\ast }\rho \left( a_{\downarrow }a_{\uparrow }\right)
+\rho \left( a_{\uparrow }^{\ast }a_{\downarrow }^{\ast }\right)
a_{\downarrow }a_{\uparrow }\right) .  \label{dh}
\end{equation}%
By (\ref{eq restrictedsimple}), (i)-(ii) straightforwardly follow. (iii) is
a direct consequence of the following computation:%
\begin{multline*}
\left[ dh\left( \rho \right) ,n_{\uparrow }n_{\downarrow }\right] =\gamma
a_{\uparrow }^{\ast }a_{\downarrow }^{\ast }\rho \left( a_{\downarrow
}a_{\uparrow }\right) \\
-\gamma \rho \left( a_{\uparrow }^{\ast }a_{\downarrow }^{\ast }\right)
a_{\downarrow }a_{\uparrow }.
\end{multline*}%
To obtain (iv), observe that 
\begin{multline*}
\left[ dh\left( \rho \right) ,a_{\downarrow }a_{\uparrow }\right] =2\left(
\mu -\lambda \right) a_{\downarrow }a_{\uparrow } \\
-\gamma \rho \left( a_{\downarrow }a_{\uparrow }\right) \left( n_{\uparrow
}+n_{\downarrow }-1\right) ,
\end{multline*}%
using again the CAR. Then, by combining this with (i), one computes that the
function $\mathfrak{z}_{t}:=\rho _{t}\left( a_{0,\downarrow }a_{0,\uparrow
}\right) $, $t\in \mathbb{R}$, satisfies the elementary ODE%
\begin{equation*}
\partial _{t}\mathfrak{z}_{t}\left( \rho \right) =i\mathrm{\nu }\left( \rho
\right) \mathfrak{z}_{t}\left( \rho \right) \ ,\qquad \mathfrak{z}_{0}\left(
\rho \right) =\rho \left( a_{\downarrow }a_{\uparrow }\right) \ ,
\end{equation*}%
from which (iv) directly follows.
\end{proof}

\noindent In the special case $\lambda =0$, i.e., without the Hubbard
interaction, Proposition \ref{lemma dynamics} reproduces the results of \cite%
[Section A]{Bona89} on the strong-coupling BCS\ model, written in that paper
as a permutation-invariant quantum-spin model. Observe also that $\mathrm{%
\kappa }(\rho ):=\left\vert \rho \left( a_{\downarrow }a_{\uparrow }\right)
\right\vert ^{2}$ is the Cooper-pair-condensate density, which, in this
situation, stays constant for all times, by Proposition \ref{lemma dynamics}
(iv).

Proposition \ref{lemma dynamics} leads to the exact dynamics of the
considered physical system prepared in a product state at initial time,
driven by the strong-coupling BCS-Hubbard Hamiltonian, in the infinite
volume limit. This set of states is still restrictive and our results \cite%
{Bru-pedra-MF-I,BruPedra-MFII,BruPedra-MFIII}, summarized in Appendix \ref%
{Mathematical Foundations}, go far beyond this simple case, by allowing us
to consider general periodic states as initial states, in contrast with all
previous results on lattice-fermion or quantum-spin systems with mean-field
interactions.

\subsection{From Product to Periodic States as Initial States}

The strong-coupling BCS-Hubbard model is clearly permutation-invariant%
\footnote{%
It is invariant under the transformation $\mathfrak{p}_{\pi }:a_{x,\mathrm{s}%
}\mapsto a_{\pi (x),\mathrm{s}}$ with $x\in \mathbb{Z}^{d}$ and $\mathrm{s}%
\in \{\uparrow ,\downarrow \}$, for all bijective mappings $\pi :\mathbb{Z}%
^{d}\rightarrow \mathbb{Z}^{d}$ which leave all but finitely many elements
invariant. See Section \ref{Permutation-Invariant Long-Range Models}.}.
First, take a permutation-invariant\footnote{%
I.e., $\rho \circ \mathfrak{p}_{\pi }=\rho $ for all bijective mappings $\pi
:\mathbb{Z}^{d}\rightarrow \mathbb{Z}^{d}$ which leave all but finitely many
elements invariant. See (\ref{permutation inv states}).} state $\rho $ as
initial state. As is explained in Appendix \ref{Permutation-Invariant State
Space}, any permutation-invariant state can be written (or approximated, to
be more precise) as a convex combination of product states (cf. the St{\o }%
rmer theorem). Thus, let $\rho _{1},\ldots ,\rho _{n}$ be $n\in \mathbb{N}$
product states and $u_{1},\ldots ,u_{n}\in \lbrack 0,1]$ such that $%
u_{1}+\cdots +u_{n}=1$, and 
\begin{equation}
\rho =\sum_{j=1}^{n}u_{j}\rho _{j}\ .  \label{finite sum}
\end{equation}%
At fixed $L\in \mathbb{N}_{0}$, we take the restriction $\rho ^{(L)}$ of $%
\rho $ to $\mathcal{B}(\mathcal{F}_{\Lambda _{L}})$, which is thus a finite
volume permutation-invariant state, like the Gibbs state (\ref{gibbs1})
associated with the strong-coupling BCS-Hubbard model. Then, in this case,
we infer from (\ref{eq restrictedsimple0})-(\ref{eq restrictedsimple}) that,
for any time $t\in \mathbb{R}$, 
\begin{equation}
\lim_{L\rightarrow \infty }\rho ^{(L)}\circ \tau
_{t}^{(L)}=\sum_{j=1}^{n}u_{j}\rho _{j}\circ \tau _{t,0}^{\mathbf{\varpi }%
(\cdot ;\rho _{j})},  \label{finite sum2}
\end{equation}%
where by a slight abuse of notation, $\mathbf{\varpi }(\cdot ;\rho )=\mathbf{%
\varpi }(\cdot ;\rho |_{\mathcal{B}(\mathcal{F}_{\{0\}})})$. Note that all
limits on states refers to the weak$^{\ast }$ topology, basically
corresponding to apply all states of (\ref{finite sum2}) on fixed elements
of $\mathcal{B}(\mathcal{F}_{\Lambda _{L}})$, $L\in \mathbb{N}_{0}$, to
perform the limit.

For general permutation-invariant states, one has to replace the finite sum (%
\ref{finite sum}) by an integral with respect to a probability measure $\mu
_{\rho }$ on the set $E_{\otimes }$ of product states in order to generalize
(\ref{finite sum2}): Formally, for any time $t\in \mathbb{R}$,%
\begin{equation}
\lim_{L\rightarrow \infty }\rho ^{(L)}\circ \tau _{t}^{(L)}=\int_{E_{\otimes
}}\hat{\rho}\circ \tau _{t,0}^{\mathbf{\varpi }(\cdot ;\hat{\rho})}\mathrm{d}%
\mu _{\rho }\left( \hat{\rho}\right) .  \label{bary}
\end{equation}%
See, e.g., (\ref{eq restricted}) for more details. As a consequence, by
combining Proposition \ref{lemma dynamics} with such a decomposition of
permutation-invariant states into product states, we obtain all dynamical
properties of the strong-coupling BCS-Hubbard model, in any
permutation-invariant initial state.

For instance, taking the state (\ref{finite sum}) and combining (\ref{finite
sum2}) with Proposition \ref{lemma dynamics} applied to the product states $%
\rho _{j}$, $j\in \{1,\ldots ,n\}$, we obtain the Cooper-field and
condensate densities: 
\begin{equation}
\rho _{t}\left( a_{0,\downarrow }a_{0,\uparrow }\right) =\sum_{j=1}^{n}u_{j}%
\sqrt{\mathrm{\kappa }(\rho _{j})}\mathrm{e}^{i(t\mathrm{\nu }\left( \rho
_{j}\right) +\theta _{\rho _{j}})}.  \label{coherence phenomena}
\end{equation}%
In particular, the Cooper-pair-condensate density defined by $\mathrm{\kappa 
}(\rho _{t}):=\left\vert \rho _{t}\left( a_{\downarrow }a_{\uparrow }\right)
\right\vert ^{2}$ at time $t\in \mathbb{R}$\ is \emph{not anymore
necessarily constant} and can have a complicated, highly non-trivial, time
evolution, in particular when $\rho $ is not a finite sum like (\ref{finite
sum}), but only the barycenter of a general probability measure on the set
of product states, see (\ref{bary}). Physically speaking, Equation (\ref%
{coherence phenomena}) expresses an \emph{interference phenomenon} on the
Cooper-field densities in each \emph{pure} state $\rho _{j}$ for $j\in
\{1,\ldots ,n\}$.

The permutation-invariant case already applies to the (weak$^{\ast }$) limit 
$\omega ^{(\infty )}$ of the Gibbs state $\omega ^{(L)}$ (\ref{gibbs1})
which is proven to exist as a permutation-invariant state $\omega ^{(\infty
)}$ because, by \cite[Theorem 6.5]{BruPedra1}, away from the
superconducting\ critical point, it is formally given by 
\begin{equation}
\omega ^{(\infty )}=\frac{1}{2\pi }\int_{0}^{2\pi }\omega ^{(\infty ,r%
\mathrm{e}^{i\theta })}\mathrm{d}\theta  \label{dec}
\end{equation}%
with $\{\mathfrak{d}=r\mathrm{e}^{i\theta },\theta \in \lbrack 0,2\pi ]\}$
being all solutions to the variational problem (\ref{var pb}) and where the
product state $\omega ^{(\infty ,\mathfrak{d})}$ is the thermodynamic limit $%
L\rightarrow \infty $ of the Gibbs state $\omega ^{(L,\mathfrak{d})}\left(
\cdot \right) $ defined by (\ref{gibbs2}). In this case, by \cite[Theorem
6.4 and\ previous discussions]{BruPedra1},%
\begin{equation}
\omega ^{(\infty ,r\mathrm{e}^{i\theta })}\left( a_{\downarrow }a_{\uparrow
}\right) =r\mathrm{e}^{i\theta }=\mathfrak{d}\ ,\qquad \theta \in \lbrack
0,2\pi ]\ ,  \label{toto1}
\end{equation}%
and if one has a superconducting phase, i.e., $r>0$, then, by \cite[Eq.
(3.3) and Theorem 6.4 (i)]{BruPedra1}, one always has the equality 
\begin{equation}
\omega ^{(\infty ,r\mathrm{e}^{i\theta })}\left( n_{\downarrow }+n_{\uparrow
}\right) =1+2\gamma ^{-1}\left( \mu -\lambda \right)  \label{toto2}
\end{equation}%
for all $\theta \in \lbrack 0,2\pi ]$. In fact, any equilibrium state is a
state in the closed convex hull of $\{\omega ^{(\infty ,r\mathrm{e}^{i\theta
})},\theta \in \lbrack 0,2\pi ]\}$. Equations (\ref{toto1})-(\ref{toto2})
imply that, for any equilibrium state $\omega $, like $\omega ^{(\infty )}$,
the frequency $\mathrm{\nu }(\omega )$, defined in Proposition \ref{lemma
dynamics} (iv), vanishes, i.e., $\mathrm{\nu }(\omega )=0$. Hence, in this
case, by Proposition \ref{lemma dynamics}, all densities are constant in
time for any equilibrium state. The same property is also true at the
superconducting\ critical point, by \cite[Theorem 6.5 (ii)]{BruPedra1}. This
is of course coherent with the well-known stationarity of equilibrium
states. For more details on equilibrium states of mean-field models, see 
\cite{BruPedra-MFVI}.

The results presented above could still have been deduced from B\'{o}na's
ones, as it is done in \cite[Section A]{Bona89} for the strong-coupling BCS\
model, for $\mathrm{H}_{L}|_{\lambda =h=0}$ to be precise. Of course, in
this case, one has to represent the lattice-fermion systems as a
permutation-invariant quantum-spin system and a permutation-invariant state
would again be required as initial state.

Using \cite{BruPedra-MFII,BruPedra-MFIII} one can easily extend this study
of the strong-coupling BCS-Hubbard model to a much larger class of initial
states: In fact, product states are only a particular case of so-called
ergodic translation-invariant\footnote{%
I.e., it is invariant, for any $x\in \mathbb{Z}^{d}$, under the
transformation $a_{x,\mathrm{s}}\mapsto a_{x+y,\mathrm{s}}$, $y\in \mathbb{Z}%
^{d}$, $\mathrm{s}\in \{\uparrow ,\downarrow \}$. See Section \ref{section
periodic states}.} states and if the initial finite volume state $\rho
^{(L)} $ is the restriction to $\mathcal{B}(\mathcal{F}_{\Lambda _{L}})$\ of
an extreme or, equivalently, ergodic translation-invariant state\footnote{%
This state acts on the CAR algebra $\mathcal{U}$ of the lattice. See (\ref%
{CAR algebra}).}, then Equation (\ref{eq restricted}) also tells us that,
for any time $t\in \mathbb{R}$,%
\begin{equation*}
\lim_{L\rightarrow \infty }\rho ^{(L)}\circ \tau _{t}^{(L)}=\rho \circ \tau
_{t,0}^{\mathbf{\varpi }(\cdot ;\rho )}
\end{equation*}%
(in the weak$^{\ast }$ sense, as before), where, again by a slight abuse of
notation, $\mathbf{\varpi }(\cdot ;\rho )=\mathbf{\varpi }(\cdot ;\rho |_{%
\mathcal{B}(\mathcal{F}_{\{0\}})})$. What's more, since 
\begin{equation*}
\tau _{t,0}^{\mathbf{\varpi }(\cdot ;\rho )}\left( \mathcal{B}\left( 
\mathcal{F}_{\{0\}}\right) \right) \subseteq \mathcal{B}\left( \mathcal{F}%
_{\{0\}}\right) \ ,
\end{equation*}%
because (\ref{Hamiltonian BCS-Hubbard approx}) is a sum of on-site terms,
the time evolution of the electron, magnetization, Coulomb correlation,
Cooper-pair-condensate and the Cooper-field densities can directly be
deduced from Proposition \ref{lemma dynamics}, for extreme (ergodic),
translation-invariant, initial states. Similar to (\ref{finite sum})-(\ref%
{finite sum2}), these quantities can be derived for general
translation-invariant states, by using their decompositions (\ref{affine
decomposition0}) in terms of extreme (or ergodic) translation-invariant
states.

All these outcomes can be extended to the case of general periodic initial
states, via straightforward modifications: for any $(\ell _{1},\ldots ,\ell
_{d})\in \mathbb{N}^{d}$ and initial $(\ell _{1},\ldots ,\ell _{d})$\emph{-}%
periodic\footnote{%
See Section \ref{section periodic states} for more details.} state $\rho $,
replace in all the above discussions on translation-invariant initial states
terms like $\rho \left( a_{\downarrow }a_{\uparrow }\right) =\rho \left(
a_{0,\downarrow }a_{0,\uparrow }\right) $ by%
\begin{equation}
\frac{1}{\ell _{1}\cdots \ell _{d}}\sum\limits_{x=(x_{1},\ldots
,x_{d}),\;x_{i}\in \{0,\ldots ,\ell _{i}-1\}}\rho \left( a_{x,\downarrow
}a_{x,\uparrow }\right) \ .  \label{ddd}
\end{equation}%
Cf. (\ref{approx interaction})-(\ref{eq:enpersite}). This goes far beyond
all previous studies on lattice-fermion or quantum-spin systems with
mean-field interactions.

\section{Entanglement of Classical and Quantum Dynamics\label{Section
Entanglement}}

Quoting \cite[p. 106]{Bokulich}, the \textquotedblleft research in
semiclassical mechanics, and especially in the subfield of quantum chaos,
has revealed that the relationship between classical and quantum mechanics
is much more subtle and intricate than the simple statement $\hbar
\rightarrow 0$ might lead us believe.\textquotedblright\ In this section, we
explicitly show an \emph{intricate} combination of classical and quantum
dynamics in mean-field systems. In order to illustrate this fact in a simple
manner, we again use our pedagogical example, that is, the strong-coupling
BCS-Hubbard model. We start by describing the classical part of the dynamics.

\subsection{Emergence of Classical Mechanics}

In the previous sections we rigorously derive the infinite volume dynamics
of the BCS-Hubbard model, which is a model comprising mean-field
interactions, and now one may ask how a classical dynamics appears in this
scope. To unveil it, first observe from Proposition \ref{lemma dynamics}
that we recover the equation of a symmetric rotor: Fix an even on-site state 
$\rho $. For any $t\in \mathbb{R}$, define the 3D vector $(\Omega
_{1}(t),\Omega _{2}(t),\Omega _{3}(t))$ by%
\begin{equation*}
\rho _{t}\left( a_{0,\downarrow }a_{0,\uparrow }\right) =\Omega
_{1}(t)+i\Omega _{2}(t)
\end{equation*}%
and 
\begin{equation}
\Omega _{3}\left( t\right) :=2\left( \mu -\lambda \right) +\gamma \left(
1-\rho _{t}\left( n_{0,\uparrow }+n_{0,\downarrow }\right) \right) .
\label{dssda}
\end{equation}%
Then, this time-dependent 3D vector satisfies, for any time $t\in \mathbb{R}$%
, the following system of ODEs:%
\begin{equation}
\left\{ 
\begin{array}{l}
\dot{\Omega}_{1}\left( t\right) =-\Omega _{3}\left( t\right) \Omega
_{2}\left( t\right) \ , \\ 
\dot{\Omega}_{2}\left( t\right) =\Omega _{3}\left( t\right) \Omega
_{1}\left( t\right) \ , \\ 
\dot{\Omega}_{3}\left( t\right) =0\ .\text{ }%
\end{array}%
\right.  \label{rotor}
\end{equation}%
It describes the time evolution of the angular momentum of a symmetric rotor
in \emph{classical} mechanics. This is \emph{not} accidental.

As a matter of fact, the equation governing the (infinite volume) mean-field
dynamics can be written in terms of Poisson brackets, i.e., as some \emph{%
Liouville's equation} of classical mechanics: In the algebraic approach to
classical mechanics \cite[Chapter 12]{quantum theory}, it is natural to
consider real- or complex-valued functions on a \emph{phase space} $\mathcal{%
P}$. Because of the self-consistency equation (\ref{self-consitency}), we
thus define a classical algebra of observables to be the real part of the
(commutative $C^{\ast }$-)algebra $C(\mathcal{P};\mathbb{C})$ of continuous
functions on the space $\mathcal{P}\equiv E_{\{0\}}^{+}$ of all even states
acting on $\mathcal{B}(\mathcal{F}_{\{0\}})$. The self-consistency equation
leads to a group\footnote{%
The fact it is a group is not that obvious, a priori. See \cite[Proposition
4.4]{Bru-pedra-MF-I} for a general proof.} $(V_{t})_{t\in \mathbb{R}}$ of
automorphisms of $C(\mathcal{P};\mathbb{C})$ defined by 
\begin{equation}
\left[ V_{t}f\right] \left( \rho \right) :=f\left( \mathbf{\varpi }(t;\rho
)\right)  \label{classical dynamics}
\end{equation}%
for any state $\rho \in \mathcal{P}$, function $f\in C(\mathcal{P};\mathbb{C}%
)$ and time $t\in \mathbb{R}$. The equation governing this dynamics can be
written in terms of Poisson brackets: \medskip

\noindent \textbf{Poisson bracket.}\emph{\ }Similar to (\ref{ddddd}), for
any $n\in \mathbb{N}$, $A_{1},\ldots ,A_{n}\in \mathcal{B}(\mathcal{F}%
_{\{0\}})$ and $g\in C^{1}\left( \mathbb{R}^{n},\mathbb{C}\right) $, we
define the function $\Gamma _{g}\in C(\mathcal{P};\mathbb{C})$ by%
\begin{equation*}
\Gamma _{g}\left( \rho \right) :=g\left( \rho \left( A_{1}\right) ,\ldots
,\rho \left( A_{n}\right) \right) ,\qquad \rho \in \mathcal{P}.
\end{equation*}%
A polynomial function in $C(\mathcal{P};\mathbb{C})$ is a function $f$ of
the form $\Gamma _{g}$ for some polynomial $g$ of $n\in \mathbb{N}$
variables. Similar to (\ref{Ynbis}), for such a function and any $\rho \in 
\mathcal{P}$, define 
\begin{multline*}
\mathrm{D}\Gamma _{g}\left( \rho \right) 
:=%
\sum_{j=1}^{n}\left( A_{j}-\rho \left( A_{j}\right) \mathbf{1}_{\mathcal{B}(%
\mathcal{F}_{\{0\}})}\right) \\
\times \partial _{x_{j}}g\left( \rho \left( A_{1}\right) ,\ldots ,\rho
\left( A_{n}\right) \right) .
\end{multline*}%
Note that $\mathrm{D}\Gamma _{g}\left( \rho \right) \in \mathcal{B}(\mathcal{%
F}_{\{0\}})$. This definition comes from a notion of convex derivative
introduced by us, as explained in \cite[Section 3.4]{Bru-pedra-MF-I}. Then,
for all functions of the form $\Gamma _{h}$ and $\Gamma _{g}$ with $g\in
C^{1}\left( \mathbb{R}^{n},\mathbb{C}\right) $ and $h\in C^{1}\left( \mathbb{%
R}^{m},\mathbb{C}\right) $ ($n,m\in \mathbb{N}$), we define the continuous
function $\left\{ \Gamma _{h},\Gamma _{g}\right\} \in C(\mathcal{P};\mathbb{C%
})$ by%
\begin{equation*}
\left\{ \Gamma _{h},\Gamma _{g}\right\} \left( \rho \right) :=\rho \left( i 
\left[ \mathrm{D}\Gamma _{h}\left( \rho \right) ,\mathrm{D}\Gamma _{g}\left(
\rho \right) \right] \right)
\end{equation*}%
for any $\rho \in \mathcal{P}$. This defines a Poisson bracket on the space
of all (local) polynomial functions on $\mathcal{P}$. See \cite[Proposition
3.11]{Bru-pedra-MF-I} for a general proof.\medskip

\noindent \textbf{Liouville's equation.}\emph{\ }The \emph{classical}
Hamiltonian $\mathrm{h}\in C(\mathcal{P};\mathbb{C})$ related to the
strong-coupling BCS-Hubbard model is a polynomial in $C(\mathcal{P};\mathbb{C%
})$ defined \emph{in a very natural way} by%
\begin{equation*}
\mathrm{h}\left( \rho \right) :=\rho \left( h\right) -\gamma \left\vert \rho
\left( a_{\uparrow }a_{\downarrow }\right) \right\vert ^{2},\quad \rho \in 
\mathcal{P},
\end{equation*}%
with $h\equiv h_{0}$ defined by (\ref{strong coupling hamb}) for $x=0$, the $%
0$ indices of operators acting on $\mathcal{F}_{\{0\}}$ having been omitted
for notational simplicity. It leads to a state-dependent Hamiltonian equal
to 
\begin{eqnarray}
\mathrm{Dh}\left( \rho \right) &=&h-\gamma \left( a_{\uparrow }^{\ast
}a_{\downarrow }^{\ast }\rho \left( a_{\downarrow }a_{\uparrow }\right)
+\rho \left( a_{\uparrow }^{\ast }a_{\downarrow }^{\ast }\right)
a_{\downarrow }a_{\uparrow }\right)  \notag \\
&&+\left( 2\gamma \left\vert \rho \left( a_{\downarrow }a_{\uparrow }\right)
\right\vert ^{2}-\rho \left( h\right) \right) \mathbf{1}_{\mathcal{B}(%
\mathcal{F}_{\{0\}})}  \notag \\
&&  \label{Dh1}
\end{eqnarray}%
for any $\rho \in \mathcal{P}$. Compare with (\ref{dh}). Then, we can
rigorously derive \emph{Liouville's equation} (see, e.g., \cite[Proposition
10.2.3]{classical-dynamics}) for any polynomial $f$ in $C(\mathcal{P};%
\mathbb{C})$:%
\begin{equation}
\partial _{t}V_{t}\left( f\right) =V_{t}\left( \{\mathrm{h},f\}\right)
=\left\{ \mathrm{h},V_{t}(f)\right\} ,\quad t\in \mathbb{R}.  \label{Dh2}
\end{equation}%
See Equation (\ref{Liouville's equations permutation}). Liouville's equation
is written here on a \emph{finite-dimensional} phase space and can easily be
studied analytically. Its solution at fixed initial state gives access to
all dynamical properties of product states driven by the strong-coupling
BCS-Hubbard model in the thermodynamic limit. In particular, it is
straightforward to check the validity of Proposition \ref{lemma dynamics}
from this equation: (i) $V_{t}\left( \mathrm{d}\right) =\mathrm{d}$; (ii) $%
V_{t}\left( \mathrm{m}\right) =\mathrm{m}$; (iii) $V_{t}\left( \mathrm{w}%
\right) =\mathrm{w}$; (iv) $V_{t}\left( \mathfrak{z}_{0}\right) =\mathfrak{z}%
_{t}$ with $\mathfrak{z}_{t}(\rho ):=\rho _{t}\left( a_{0,\downarrow
}a_{0,\uparrow }\right) $ for $\rho \in \mathcal{P}$ and $t\in \mathbb{R}$.

The time evolution $V_{t}\left( \mathrm{p}_{n}\right) $ of the \emph{%
non-affine} polynomials%
\begin{eqnarray*}
&&\mathrm{p}_{n}(\rho )%
:=%
\left\vert \rho \left( a_{\downarrow }a_{\uparrow }\right) \right\vert ^{2n}=%
\frac{1}{4^{n}}(\rho \left( a_{\downarrow }a_{\uparrow }+a_{\uparrow }^{\ast
}a_{\downarrow }^{\ast }\right) ^{2} \\
&&+\rho \left( ia_{\downarrow }a_{\uparrow }-ia_{\uparrow }^{\ast
}a_{\downarrow }^{\ast }\right) ^{2})^{n},
\end{eqnarray*}%
$\rho \in \mathcal{P}$, for any integer $n\geq 1$ can be obtained by using (%
\ref{Dh2}). In particular, for $n=1$, since the (convex) derivative of $%
\mathrm{p}_{1}$ at $\rho \in \mathcal{P}$ equals 
\begin{multline*}
\mathrm{Dp}_{1}\left( \rho \right) =a_{\uparrow }^{\ast }a_{\downarrow
}^{\ast }\rho \left( a_{\downarrow }a_{\uparrow }\right) +\rho \left(
a_{\uparrow }^{\ast }a_{\downarrow }^{\ast }\right) a_{\downarrow
}a_{\uparrow } \\
-2\left\vert \rho \left( a_{\downarrow }a_{\uparrow }\right) \right\vert ^{2}%
\mathbf{1}_{\mathcal{B}(\mathcal{F}_{\{0\}})},
\end{multline*}%
one directly recover from Liouville's equation, combined with the CAR and (%
\ref{Dh1}), that the Cooper-pair-condensate density is static. Compare with
Proposition \ref{lemma dynamics} (iv). Moreover, by considering
complex-valued polynomials $g$ in the space%
\begin{multline*}
\left\{ \left( x,y\right) \in \mathbb{R}^{2}:x^{2}+y^{2}\leq 1\right\} \\
\times \lbrack 2\left( \mu -\lambda \right) -\gamma ,2\left( \mu -\lambda
\right) +\gamma ],
\end{multline*}%
of $(\Omega _{1},\Omega _{2},\Omega _{3})$-coordinates one can recover the
classical dynamics of a symmetric rotor, as stated in (\ref{rotor}). In
fact, one can write a(nother) Liouville's equation on a convenient reduced
(or effective) phase space. The real and imaginary parts of $\rho \left(
a_{\downarrow }a_{\uparrow }\right) $ (Cooper-field densities), respectively 
$\Omega _{1}$ and $\Omega _{2}$, and the shifted particle density $\Omega
_{3}$ (\ref{dssda}) represent three physical quantities that can be seen as 
\emph{macroscopic} in the fermionic system under consideration. See, e.g., (%
\ref{dynamics approx0}).

To conclude this subsection, recall that the classical dynamics presented
above has the space $\mathcal{P}\equiv E_{\{0\}}^{+}$ of all even states
acting on $\mathcal{B}(\mathcal{F}_{\{0\}})$ as phase space, i.e., this
dynamics is defined on $C(\mathcal{P};\mathbb{C})$. Taking a broader
perspective, a classical dynamics can also be defined on $C(E_{\Pi };\mathbb{%
C})$, with $E_{\Pi }$ being the space of permutation-invariant states on $%
\mathcal{U}$, the CAR algebra of the infinite lattice. For more details, see
Appendix \ref{sect Permutation-Invariant Long-Range Models}. In this case,
the classical dynamics constructed on $C(E_{\Pi };\mathcal{U})$ can be
pushed forward, through the restriction mapping $E_{\Pi }\rightarrow 
\mathcal{P}$, from $C(E_{\Pi };\mathbb{C})$ to $C(\mathcal{P};\mathbb{C})$.
For an even more general definition of classical dynamics, which can be
extended to periodic states, see Appendix \ref{Section classical
dynamics-general}.

\subsection{Classical Versus Quantum Pictures}

For product states at initial time, in the case of the strong-coupling
BCS-Hubbard model, it is natural to restrict the quantum observables to the
algebra $\mathcal{B}(\mathcal{F}_{\{0\}})$ of linear operators on the
fermion Fock space $\mathcal{F}_{\{0\}}$. This keeps things simple. In this
case, for any even state $\rho $ on $\mathcal{B}(\mathcal{F}_{\{0\}})$, we
can define a non-autonomous quantum dynamics by the continuous evolution%
\footnote{%
It satisfies the reverse cocycle property: $\tau _{t,s}^{(\mathbf{\varpi }%
(\cdot ;\rho ))}=\tau _{r,s}^{(\mathbf{\varpi }(\cdot ;\rho ))}\circ \tau
_{t,r}^{(\mathbf{\varpi }(\cdot ;\rho ))}$ for any $s,r,t\in \mathbb{R}$.}
family $(\tau _{t,s}^{(\mathbf{\varpi }(\cdot ;\rho ))})_{_{s,t\in \mathbb{R}%
}}$ of automorphisms of $\mathcal{B}(\mathcal{F}_{\{0\}})$, defined by (\ref%
{idiot}) for $\omega =\mathbf{\varpi }(\cdot ;\rho )$. The physical
relevance of this dynamics comes from Equations (\ref{eq restrictedsimple}).
Therefore, for initial product states and on on-site observables, the
mean-field dynamics can be seen either as a classical one on $C(\mathcal{P};%
\mathbb{C})$ or as a non-autonomous quantum dynamics on $\mathcal{B}(%
\mathcal{F}_{\{0\}})$. The classical dynamics uniquely defines the quantum
dynamics and conversely.

For initial states that are not product states, the situation is more
involved, but also much more interesting, since interference phenomena on
macroscopic quantities may take place. See, e.g., (\ref{coherence phenomena}%
).

Let us consider general permutation-invariant states (i.e., not necessarily
product states) as initial states. In this case, the quantum world refers to 
\emph{all} local observables of the infinite lattice and we thus have to
consider the CAR $C^{\ast }$-algebra 
\begin{equation}
\mathcal{U}\varsupsetneq \bigcup_{L\in \mathbb{N}}\mathcal{B}(\mathcal{F}%
_{\Lambda _{L}})\varsupsetneq \mathcal{B}(\mathcal{F}_{\{0\}}),
\label{CAR algebra}
\end{equation}%
which is the $C^{\ast }$-algebra generated by all finite volume quantum
observables for fermions in the lattice. See Appendix \ref{Algebra of
Lattices}. This algebra corresponds to what we call the \textquotedblleft 
\textit{primordial}\textquotedblright\ quantum algebra in our general
abstract setting, introduced in \cite{Bru-pedra-MF-I}

Still in relation to the terminology we introduce in \cite{Bru-pedra-MF-I},
the \textit{secondary} quantum algebra corresponds here to the $C^{\ast }$%
-algebra $C(E_{\Pi };\mathcal{U})$ of all continuous $\mathcal{U}$-valued
functions on the space $E_{\Pi }$ of permutation-invariant states on $%
\mathcal{U}$. This is nothing else\footnote{%
Up to an isomorphism. See \cite[Section 1]{Bru-pedra-MF-I} for very general
mathematical arguments proving that fact.} than the following tensor
product: 
\begin{equation}
C(E_{\Pi };\mathcal{U})\equiv C(E_{\Pi };\mathbb{C})\otimes \mathcal{U}.
\label{fffff}
\end{equation}%
Having in mind that a classical dynamics can be defined on $C(E_{\Pi };%
\mathbb{C})$, this is similar to quantum-classical hybrid theories of
theoretical physics, described for instance in \cite%
{extra-ref1988,extra-ref,extra-ref2,extra-ref3,extra-ref00,extra-ref1}. With
this definition we naturally have the inclusions $\mathcal{U}\subseteq
C(E_{\Pi };\mathcal{U})\ $and$\ C(E_{\Pi };\mathbb{C})\subseteq C(E_{\Pi };%
\mathcal{U})$, by identifying elements of $\mathcal{U}$ with constant
functions and elements of $C(E_{\Pi };\mathbb{C})$ with functions whose
values are scalar multiples of the unit of the primordial algebra $\mathcal{U%
}$.

The quantum (short-range) dynamics on the secondary quantum algebra $%
C(E_{\Pi };\mathcal{U})$ refers to the continuous evolution\footnote{%
It satisfies the reverse cocycle property: $\mathfrak{T}_{t,s}=\mathfrak{T}%
_{r,s}\circ \mathfrak{T}_{t,r}$ for any $s,r,t\in \mathbb{R}$.} family $(%
\mathfrak{T}_{t,s})_{s,t\in \mathbb{R}}$ of $\ast $-automorphisms\footnote{%
The mathematical fact that it is a continuous evolution family of
automorphisms is not obvious, a priori. The proof uses that $E_{\Pi }$ is a
metrizable weak$^{\ast }$-compact space, by separability of $\mathcal{U}$.
See \cite[Lemma 5.2]{Bru-pedra-MF-I} for a general proof.} of $C(E_{\Pi };%
\mathcal{U})$ defined by%
\begin{equation*}
\left[ \mathfrak{T}_{t}\left( f\right) \right] \left( \rho \right) :=\tau
_{t,0}^{\mathbf{\varpi }(\cdot ;\rho )}\left( f\left( \rho \right) \right)
,\qquad \rho \in E_{\Pi },
\end{equation*}%
for any function $f\in C(E_{\Pi };\mathcal{U})$ and time $t\in \mathbb{R}$,
where, again by a slight abuse of notation, $\mathbf{\varpi }(\cdot ;\rho )=%
\mathbf{\varpi }(\cdot ;\rho |_{\mathcal{B}(\mathcal{F}_{\{0\}})})$. This
state-dependent dynamics lets every element of the classical algebra $%
C(E_{\Pi };\mathbb{C})$ invariant, i.e., $\mathfrak{T}_{t}\left( f\right) =f$
for any classical function $f\in C(E_{\Pi };\mathbb{C})$. In other words,
the classical algebra $C(E_{\Pi };\mathbb{C})$ is a subalgebra of the
so-called fix point algebra of the family $(\mathfrak{T}_{t})_{t\in \mathbb{R%
}}$ of $\ast $-automorphisms of $C(E_{\Pi };\mathcal{U})$.

The physical relevance of the above mathematical structure comes from the
fact that, for each time $t\in \mathbb{R}$, permutation-invariant state $%
\rho \in E_{\Pi }$ and any element $A\in \mathcal{U}\subseteq C(E_{\Pi };%
\mathcal{U})$, 
\begin{multline*}
\lim_{L\rightarrow \infty }\rho ^{(L)}\circ \tau _{t}^{(L)}\left( A\right)
=\int_{E_{\otimes }}\hat{\rho}\circ \tau _{t,0}^{\mathbf{\varpi }(\cdot ;%
\hat{\rho})}\left( A\right) \mathrm{d}\mu _{\rho }\left( \hat{\rho}\right) \\
=\rho \circ \mathfrak{T}_{t}\left( A\right) ,
\end{multline*}%
by Equation (\ref{bary}), where in the last equality $\rho $ is seen as a
state of $C(E_{\Pi };\mathcal{U})$ via the definition%
\begin{equation*}
\rho \left( f\right) \doteq \int_{E_{\otimes }}\hat{\rho}\left( f\left( \hat{%
\rho}\right) \right) \mathrm{d}\mu _{\rho }\left( \hat{\rho}\right)
\end{equation*}%
for any $f\in C(E_{\Pi };\mathcal{U})$. See \cite[Theorem 4.3]%
{BruPedra-MFIII} for the general mathematical statement. The classical part
of the full mean-field dynamics explicitly appears in the time evolution of
product states $\hat{\rho}\in E_{\otimes }$ (cf. the St{\o }rmer theorem)
and is related to a Liouville's equation in the classical (i.e.,
commutative) algebra of continuous functions $C(\mathcal{P};\mathbb{C})$, as
explained in the previous subsection.

In the theoretical framework we present here, the classical and quantum
worlds are intrinsically interdependent, in the following manner:

\begin{itemize}
\item The quantum (short-range) dynamics on $C(E_{\Pi };\mathcal{U})$ yields
a well-defined classical dynamics on $C(\mathcal{P};\mathbb{C})$, by
restriction on product states.

\item Conversely, the classical dynamics on $C(\mathcal{P};\mathbb{C})$
uniquely defines a quantum (short-range) dynamics on $C(E_{\Pi };\mathcal{U}%
) $.
\end{itemize}

\noindent This is a mathematical fact proven for general quantum systems in 
\cite[Sections 4.2-4.3 and 5.2]{Bru-pedra-MF-I}.

On the one hand, the classical world, represented by the commutative
(sub)algebra $C(E_{\Pi };\mathbb{C})$, is embedded in the quantum world, as
mathematically expressed by the above inclusion $C(E_{\Pi };\mathbb{C}%
)\subseteq C(E_{\Pi };\mathcal{U})$. On the other hand, our theory entangles
the quantum and classical worlds through self-consistency. As a result,
(effective) non-autonomous short-range dynamics can emerge. Seeing both
entangled worlds, quantum and classical, as \textquotedblleft two sides of
the same coin\textquotedblright\ looks like an oxymoron, but there is no
contradiction there, for everything refers to a \emph{primordial} quantum
world mathematically encoded in the structure of $\mathcal{U}$. For
instance, the phase space $\mathcal{P}$ and state space $E_{\Pi }$ are
imprints left by $\mathcal{U}\varsupsetneq \mathcal{B}(\mathcal{F}_{\{0\}})$
in the classical world, see (\ref{fffff}).

Note that if $\mathcal{U}$ was a commutative algebra, the corresponding
Poisson bracket and, hence, the dynamics would have been trivial. Observe
also that, if the primordial algebra would be $\mathcal{B}(\mathcal{F}%
_{\{0\}})$, instead of $\mathcal{U}$, then all the above construction would
be still relevant for the case of initial states being product states. In
this situation, the introduction of the secondary quantum algebra is
superfluous to derive the mean-field dynamics, whereas it becomes essential
when the initial state is permutation-invariant, but \emph{not} a product
state.

All the above construction can be extended to periodic states and general
lattice-fermion or quantum spin systems. For more details, see Appendix \ref%
{Entanglement}.

\section{Conclusions}

The dynamics of the strong-coupling BCS-Hubbard model has been exactly
derived, in the infinite volume limit. It explicitly determines, among other
things, the dynamical impact of the (screened) Coulomb repulsion on ($s$%
-wave) superconductivity. For non-pure phases, we also prove that the
Cooper-pair-condensate density is not anymore necessarily constant in time
and can have a complicated time evolution, as a consequence of interference
phenomena.

Much more importantly, this model illustrates the general behavior of
mean-field dynamics at infinite volume, as rigorously explained in Appendix %
\ref{Mathematical Foundations}. We demonstrate via this example that a
classical mechanics does not only appear in the limit $\hbar \rightarrow 0$,
as explained for instance in \cite{landsmann07,extra-ref0000}. This was
already observed by various mathematical physicists. In particular, B\'{o}%
na's major conceptual contribution \cite{Bono2000} is to highlight the
emergence of classical mechanics without necessarily the disappearance of
the quantum world. However, we propose here a \emph{new} method to
mathematically implement it, with a \emph{broader} domain of applicability
than B\'{o}na's original version \cite{Bono2000} (see also \cite%
{landsmann07,Odzijewicz,Bona2020} and references therein). For detailed
explanations, see \cite[Section 3]{Bru-pedra-MF-I}.

In contrast with \emph{all} previous approaches, including those of
theoretical physics (see, e.g., \cite%
{extra-ref1988,extra-ref,extra-ref2,extra-ref3,extra-ref00,extra-ref1,extra-ref0000,extra-ref0bis}%
), in ours the classical and quantum worlds are \emph{entangled}, with \emph{%
backreaction}\footnote{%
We do not mean here the so-called \emph{quantum backreaction}, commonly used
in physics, which refers to the backreaction effect of quantum fluctuations
on the classical degrees of freedom. Note further that the phase spaces we
consider are, generally, much more complex than those related to the
position and momentum of simple classical particles.
\par
{}} (that is, feedbacks), as expected. Differently from B\'{o}na's setting,
our perspective has the advantage to highlight inherent \emph{%
self-consistency} aspects, which are absolutely not exploited in \cite%
{Bono2000}, as well as in quantum-classical hybrid theories of physics
described, for instance, in \cite%
{extra-ref1988,extra-ref,extra-ref2,extra-ref3,extra-ref00,extra-ref1,extra-ref0,extra-ref0bis}%
.

Remark that the theoretical construction presented here is not useful when
the macroscopic\footnote{%
\textquotedblleft Macroscopic\textquotedblright\ can still mean short (even
atomic) length scales. For lattice systems, it should quantitatively be
measured in terms of lattice units (l.u.), which is typically a few \aa ngstr%
\"{o}ms. For instance, a length $L\simeq 1000$ refers to a few hundreds of
nanometers, only. One thousand is a priori a large number, but everything
depends of course on the rate of convergence of microscopic dynamics in the
thermodynamic limit $L\rightarrow \infty $. In general, this may be an
exponential rate (with respect to the volume $|\Lambda _{L}|$), similar to
what is proven in \cite{LDP,BPA2} for electric current densities in
non-interacting lattice fermions with disorder.} time evolution in the
Heisenberg picture is \emph{not} state-dependent, as in quantum lattice
systems with \emph{short-range} interactions. Nevertheless, quantum
many-body systems in the continuum are expected to have, in general, only a
state-dependent Heisenberg dynamics, as explained for instance in \cite[%
Section 6.3]{BrattelliRobinson}. Additionally, we show that such a
mathematical framework is generally \emph{imperative} to describe the
macroscopic dynamics of quantum many-body systems with mean-field
interactions, because of the necessity of coupled quantum-classical
evolution equations, implementing self-consistency \emph{when long-range
order take place}. The phenomenological aspects of quantum dynamics in
presence of mean-filed interactions discussed here and that are highlighted
by our original approach to this problem, are very likely not restricted
mean-field case only, but should also appear in presence of interactions
that are sufficiently long-range\footnote{%
In fact, the existence of long-range order in quantum systems with
sufficiently long-range interactions can be mathematically proven by using
the celebrated Bishop-Phelps theorem.} to yield non-vanishing\footnote{%
In a given representation of the observable algebra, which is fixed by the
initial state.} background fields, in the spirit of the Higgs mechanism of
quantum field theory. We therefore think that our mathematical framework for
long-range dynamics, outlined here by means of a pedagogical explicit
example, opens new theoretical perspectives\footnote{%
Even after a few centuries, the Newtonian gravitational constant is still 
\emph{not} accurately known, in comparison with all other fundamental
constants. See \cite{Naturegravity} for an account of recent experiment. It
is also very difficult to detect gravity at scales below micrometers, still
a macroscopic scale as compared with atomic ones. On the other hand,
interference phenomena for gravitational waves appear (2017 Nobel Prize in
Physics). Gravitation looks like a macroscopic background (Higgs-like) field
(cf. Bogoliubov approximation), similar to the Cooper-field densities in the
strong-coupling BCS-Hubbard model on which classical mechanics applies.} in
the understanding of the classical word within the quantum one.

\appendix%

\section{Mathematical Foundations\label{Mathematical Foundations}}

The entanglement of classical and quantum short-range dynamics in mean-field
systems refers to results obtained in \cite{BruPedra-MFII,BruPedra-MFIII} on
the dynamics of quantum lattice systems with mean-field interactions. They
are far more general than previous ones because the invariance under
permutations of lattice sites is \emph{not} required anymore:

\begin{itemize}
\item The short-range part of the corresponding Hamiltonian is very general
since only a sufficiently strong polynomial decay of its interactions and a
translation invariance are necessary.

\item The mean-field part is also very general, being an infinite sum (over $%
n$) of mean-field terms of order $n\in \mathbb{N}$. In fact, even for
permutation-invariant systems, the class of mean-field interactions we are
able to handle is much larger than what was previously studied.

\item The initial state is only required to be periodic. By \cite[%
Proposition 2.3]{BruPedra-MFII}, observe that the set of all such initial
states is dense within the set of all even states, the physically relevant
ones.
\end{itemize}

The papers \cite{BruPedra-MFII,BruPedra-MFIII} are altogether about 126
pages long. Therefore, the aim of the appendix is to present, in a concise
way, their key points, being meanwhile mathematically rigorous. Note,
however, that the contents of Appendix \ref{sect Permutation-Invariant
Long-Range Models} are new and cannot be found in \cite%
{BruPedra-MFII,BruPedra-MFIII,Bru-pedra-proceeding,BruPedra-MFVI,BruPedra1}.

\subsection{$C^{\ast }$-Algebraic Setting}

\subsubsection{CAR Algebra of Lattices\label{Algebra of Lattices}}

Let $\mathbb{Z}^{d}$ be the $d$-dimensional cubic lattice and $\mathcal{P}%
_{f}\subseteq 2^{\mathbb{Z}^{d}}$ the set of all non-empty finite subsets of 
$\mathbb{Z}^{d}$. In order to define the thermodynamic limit, for
simplicity, we use cubic boxes 
\begin{equation}
\Lambda _{L}:=\{\mathbb{Z}\cap \left[ -L,L\right] \}^{d}\ ,\qquad L\in 
\mathbb{N}_{0}\ .  \label{eq:def lambda n}
\end{equation}

Let $\mathrm{S}$ be a fixed (once and for all) finite set (of orthonormal
spin modes). For any $\Lambda \in \mathcal{P}_{f}\cup \{\mathbb{Z}^{d}\}$, $%
\mathcal{U}_{\Lambda }$ is the universal unital $C^{\ast }$-algebra\footnote{%
$\mathcal{U}_{\Lambda }\equiv \mathcal{B}(\mathbb{C}^{2^{\Lambda \times 
\mathrm{S}}})$ is equivalent to the algebra of$\mathcal{\ }2^{\left\vert
\Lambda \times \mathrm{S}\right\vert }\times 2^{\left\vert \Lambda \times 
\mathrm{S}\right\vert }$ complex matrices, when $\Lambda \in \mathcal{P}_{f}$%
.} generated by the elements $\{a_{x,\mathrm{s}}\}_{x\in \Lambda ,\mathrm{s}%
\in \mathrm{S}}$ satisfying the canonical anti-commutation relations (CAR):
for any $x,y\in \mathbb{Z}^{d}$ and $\mathrm{s},\mathrm{t}\in \mathrm{S}$,%
\begin{equation}
a_{x,\mathrm{s}}a_{y,\mathrm{t}}+a_{y,\mathrm{t}}a_{x,\mathrm{s}}=0,\ a_{x,%
\mathrm{s}}a_{y,\mathrm{t}}^{\ast }+a_{y,\mathrm{t}}^{\ast }a_{x,\mathrm{s}%
}=\delta _{\mathrm{s},\mathrm{t}}\delta _{x,y}\mathfrak{1}.  \label{CARbis}
\end{equation}%
Here, $\delta _{k,l}$ is the Kronecker delta, that is, the function of two
variables defined by $\delta _{k,l}:=1$ if $k=l$ and $\delta _{k,l}=0$
otherwise. Note that we use the notation $\mathcal{U}\equiv \mathcal{U}_{%
\mathbb{Z}^{d}}$ and define 
\begin{equation}
\mathcal{U}_{0}:=\bigcup_{\Lambda \in \mathcal{P}_{f}}\mathcal{U}_{\Lambda },
\label{simple}
\end{equation}%
which is a dense normed $\ast $-subalgebra of $\mathcal{U}$. In particular, $%
\mathcal{U}$\ is separable, since, for every finite region $\Lambda \in 
\mathcal{P}_{f}$, $\mathcal{U}_{\Lambda }$ has finite dimension. Elements of 
$\mathcal{U}_{0}$ are called local elements. The (real) Banach subspace of
all self-adjoint elements of $\mathcal{U}$ is denoted by $\mathcal{U}^{%
\mathbb{R}}\varsubsetneq \mathcal{U}$.

Translations are represented by a group homomorphism $x\mapsto \alpha _{x}$
from $\mathbb{Z}^{d}$ to the group of $\ast $-automorphisms of $\mathcal{U}$%
, which is uniquely defined by the condition%
\begin{equation}
\alpha _{x}(a_{y,\mathrm{s}})=a_{y+x,\mathrm{s}},\quad y\in \mathbb{Z}^{d},\;%
\mathrm{s}\in \mathrm{S}.  \label{transl}
\end{equation}%
The mapping $x\mapsto \alpha _{x}$ is used below to define symmetry groups
of states as well as translation-invariant interactions of lattice-fermion
systems.

The results presented in the current paper also hold true in the context of
quantum-spin systems, but we focus on lattice-fermion systems which are,
from a technical point of view, slightly more difficult. In fact, the
additional difficulty in Fermi systems is that, for any disjoint $\Lambda
^{(1)},\Lambda ^{(2)}\in \mathcal{P}_{f}$ and elements $B_{1}\in \mathcal{U}%
_{\Lambda ^{(1)}}$, $B_{2}\in \mathcal{U}_{\Lambda ^{(2)}}$, the commutator 
\begin{equation*}
\left[ B_{1},B_{2}\right] :=B_{1}B_{2}-B_{2}B_{1}
\end{equation*}%
may be non-zero, in general. For instance, the CAR (\ref{CARbis}) trivially
yield $[a_{x,\mathrm{s}},a_{y,\mathrm{t}}]=2a_{x,\mathrm{s}}a_{y,\mathrm{t}%
}\neq 0$ for any $x,y\in \mathfrak{L}$ and $\mathrm{s},\mathrm{t}\in \mathrm{%
S}$, $(x,\mathrm{s})\neq (y,\mathrm{t})$. Because of the CAR (\ref{CARbis}),
the commutation property of disjoint lattice regions is satisfied for all 
\emph{even} elements, which are defined as follows: The condition 
\begin{equation}
\sigma (a_{x,\mathrm{s}})=-a_{x,\mathrm{s}},\qquad x\in \Lambda ,\ \mathrm{s}%
\in \mathrm{S}\ ,  \label{automorphism gauge invariance}
\end{equation}%
defines a unique $\ast $-automorphism $\sigma $ of the $C^{\ast }$-algebra $%
\mathcal{U}$. The subspace 
\begin{equation}
\mathcal{U}^{+}:=\{A\in \mathcal{U}:A=\sigma (A)\}
\label{definition of even operators}
\end{equation}%
is the $C^{\ast }$-subalgebra of so-called even elements of $\mathcal{U}$.
Then, for any disjoint $\Lambda ^{(1)},\Lambda ^{(2)}\in \mathcal{P}_{f}$, 
\begin{equation*}
\left[ B_{1},B_{2}\right] =0,\qquad B_{1}\in \mathcal{U}_{\Lambda
^{(1)}}\cap \mathcal{U}^{+},\ B_{2}\in \mathcal{U}_{\Lambda ^{(2)}}.
\end{equation*}

This last condition is the expression of the principle of locality in
quantum field theory. Using well-known constructions\footnote{%
More precisely, the so-called sector theory of quantum field theory.}, the $%
C^{\ast }$-algebra $\mathcal{U}$, generated by anticommuting elements, can
be recovered from $\mathcal{U}^{+}$. As a consequence, the $C^{\ast }$%
-algebra $\mathcal{U}^{+}$ should thus be seen as more fundamental than $%
\mathcal{U}$ in Physics. In fact, $\mathcal{U}$ corresponds in this context
to the so-called local field algebra. See, e.g., \cite[Sections 4.8 and 6]%
{Araki-livre}.

The principle of locality of quantum field theory, usually invoked to see
the algebra $\mathcal{U}^{+}$ as being more fundamental than $\mathcal{U}$,
does not prevent us from considering mean-field interactions as possibly
fundamental interactions, as explained in\ \cite[Section 1]{BruPedra-MFIII}.
The choice of $\mathcal{U}^{+}$ only refers to the fact that measurable
physical quantities (observables) are local. By contrast, the full energy of
lattice-fermion systems with short-range and mean-field interactions are
generally inaccessible to measurements,\ in infinite volume. In fact, the
mean-field part yields possibly non-vanishing (effective) background fields,
in the spirit of the Higgs mechanism of quantum field theory, in a given
representation of the observable algebra, which is fixed by the initial
state.

\subsubsection{Short-Range Interactions\label{Short-Range Interactions}}

A (complex) \emph{interaction} is a mapping $\Phi :\mathcal{P}%
_{f}\rightarrow \mathcal{U}^{+}$ such that $\Phi _{\Lambda }\in \mathcal{U}%
_{\Lambda }$ for all $\Lambda \in \mathcal{P}_{f}$. The set of all
interactions can be naturally endowed with the structure of a complex vector
space.\ By using the norm 
\begin{equation}
\left\Vert \Phi \right\Vert _{\mathcal{W}}%
:=%
\underset{x,y\in \mathbb{Z}^{d}}{\sup }\sum\limits_{\Lambda \in \mathcal{P}%
_{f},\;\Lambda \supseteq \{x,y\}}\frac{\Vert \Phi _{\Lambda }\Vert _{%
\mathcal{U}}}{\mathbf{F}\left( x,y\right) },  \label{iteration00}
\end{equation}%
where, given some fixed parameters $\epsilon ,\varsigma >0$, for any $x,y\in 
\mathbb{Z}^{d}$, 
\begin{equation}
\mathbf{F}\left( x,y\right) 
:=%
\mathrm{e}^{2\varsigma \left\vert x-y\right\vert }\left( 1+\left\vert
x-y\right\vert \right) ^{(d+\epsilon )}.  \label{iteration01}
\end{equation}%
We then define a separable Banach space $\mathcal{W}$ of short-range
interactions, which are, by definition, those interactions that have finite
norm. Here, $\left\vert \cdot -\cdot \right\vert $ stands for the Euclidean
metric. The particular choice of function (\ref{iteration01}) defining the
norm (\ref{iteration00}) is made only for simplicity and a much more general
class of space decays could be considered, as discussed in \cite[Section 3.1]%
{BruPedra-MFII}. We use in the sequel three important properties of
short-range interactions:\bigskip 

\noindent \underline{(i) Self-adjointness:} There is a natural involution $%
\Phi \mapsto \Phi ^{\ast }:=(\Phi _{\Lambda }^{\ast })_{\Lambda \in \mathcal{%
P}_{f}}$ defined on the Banach space $\mathcal{W}$ of short-range
interactions. Self-adjoint interactions are, by definition, interactions $%
\Phi $ satisfying $\Phi =\Phi ^{\ast }$. The (real) Banach subspace of all
self-adjoint short-range interactions is denoted by $\mathcal{W}^{\mathbb{R}%
}\varsubsetneq \mathcal{W}$, similar to $\mathcal{U}^{\mathbb{R}%
}\varsubsetneq \mathcal{U}$.\bigskip

\noindent \underline{(ii) Translation invariance:} By definition, the
interaction $\Phi $ is translation-invariant\ if%
\begin{equation*}
\Phi _{\Lambda +x}=\alpha _{x}\left( \Phi _{\Lambda }\right) ,\qquad x\in 
\mathbb{Z}^{d},\ \Lambda \in \mathcal{P}_{f},
\end{equation*}%
where%
\begin{equation*}
\Lambda +x:=\left\{ y+x\in \mathbb{Z}^{d}:y\in \Lambda \right\} .
\end{equation*}%
Here, $\{\alpha _{x}\}_{x\in \mathbb{Z}^{d}}$ is the family of (translation) 
$\ast $-automorphisms of $\mathcal{U}$ defined by (\ref{transl}). We then
denote by $\mathcal{W}_{1}\varsubsetneq \mathcal{W}$ the (separable) Banach
subspace of translation-invariant, short-range interactions on $\mathbb{Z}%
^{d}$. \bigskip

\noindent \underline{(iii) Finite range:} For any $\mathrm{R}\in \left[
0,\infty \right) $, we define the closed subspace%
\begin{eqnarray}
\mathcal{W}^{\mathrm{R}}%
:=%
\Big\{%
\Phi &\in &\mathcal{W}_{1}:\Phi _{\Lambda }=0\text{ for }\Lambda \in 
\mathcal{P}_{f}  \label{finite range} \\
&&\left. \text{such that }\max\limits_{x,y\in \Lambda }\{\left\vert
x-y\right\vert \}>\mathrm{R}\right. 
\Big\}
\notag
\end{eqnarray}%
of finite-range, translation-invariant interactions. For $\mathrm{R}=0$, we
obtain the space $\mathcal{W}_{\Pi }:=\mathcal{W}^{0}$ of
permutation-invariant interactions described in Appendix \ref{sect
Permutation-Invariant Long-Range Models}. \bigskip

Short-range interactions define sequences of local (complex) energy
elements: For any $\Phi \in \mathcal{W}$ and $L\in \mathbb{N}_{0}$,%
\begin{equation}
U_{L}^{\Phi }:=\sum\limits_{\Lambda \subseteq \Lambda _{L}}\Phi _{\Lambda
}\in \mathcal{U}_{\Lambda _{L}}\cap \mathcal{U}^{+},  \label{local energy}
\end{equation}%
where we recall that $\Lambda _{L}:=\{\mathbb{Z}\cap (-L,L)\}^{d}$ is the
cubic box used to define the thermodynamic limit (see (\ref{eq:def lambda n}%
)). The energy elements $U_{L}^{\Phi }$, $L\in \mathbb{N}_{0}$, refer to an
extensive quantity since their norm are proportional to the volume of the
region they correspond to: In fact, for any $L\in \mathbb{N}_{0}$ and $\Phi
\in \mathcal{W}$, 
\begin{equation}
\left\Vert U_{L}^{\Phi }\right\Vert _{\mathcal{U}}\leq \mathbf{C}\left\vert
\Lambda _{L}\right\vert \left\Vert \Phi \right\Vert _{\mathcal{W}},
\label{norm Uphi}
\end{equation}%
where 
\begin{equation}
\mathbf{C}:=\sum_{x\in \mathbb{Z}^{d}}\frac{1}{\left( 1+\left\vert
x\right\vert \right) ^{d+\epsilon }}<\infty .  \label{constant}
\end{equation}%
Moreover, for any self-adjoint interaction $\Phi \in \mathcal{W}^{\mathbb{R}%
} $, $U_{L}^{\Phi }\in \mathcal{U}^{\mathbb{R}}$ (i.e., $U_{L}^{\Phi
}=(U_{L}^{\Phi })^{\ast }$), $L\in \mathbb{N}_{0}$, is a sequence of local
Hamiltonians.

Each local Hamiltonian associated with $\Phi \in \mathcal{W}^{\mathbb{R}}$
leads to a local dynamics on the full $C^{\ast }$-algebra $\mathcal{U}$ via
the group $(\tau _{t}^{(L,\Phi )})_{t\in {\mathbb{R}}}$ of $\ast $-auto%
\-%
morphisms of $\mathcal{U}$ defined by 
\begin{equation}
\tau _{t}^{(L,\Phi )}(A)=\mathrm{e}^{itU_{L}^{\Phi }}A\mathrm{e}%
^{-itU_{L}^{\Phi }},\qquad A\in \mathcal{U}.  \label{quantum dyn local}
\end{equation}%
It is the continuous group which is the solution to the evolution equation%
\begin{equation*}
\forall t\in {\mathbb{R}}:\qquad \partial _{t}\tau _{t}^{(L,\Phi )}=\tau
_{t}^{(L,\Phi )}\circ \delta _{L}^{\Phi },
\end{equation*}%
where $\tau _{0}^{(L,\Phi )}=\mathbf{1}_{\mathcal{U}}$ is the identity
mapping on $\mathcal{U}$. Here, at each $L\in \mathbb{N}_{0}$ and $\Phi \in 
\mathcal{W}^{\mathbb{R}}$, $\delta _{L}^{\Phi }$ is defined on $\mathcal{U}$
by%
\begin{equation*}
\delta _{L}^{\Phi }(A):=i\left[ U_{L}^{\Phi },A\right] :=i\left( U_{L}^{\Phi
}A-AU_{L}^{\Phi }\right)
\end{equation*}%
for any $A\in \mathcal{U}$. This corresponds to a quantum dynamics in the
Heisenberg picture. Note that, for every $L\in \mathbb{N}_{0}$ and $\Phi \in 
\mathcal{W}^{\mathbb{R}}$, $\delta _{L}^{\Phi }$ is a so-called symmetric
derivation which belongs to the Banach space $\mathcal{B}(\mathcal{U})$ of
bounded operators acting on the $C^{\ast }$-algebra $\mathcal{U}$, see,
e.g., \cite[Section 3.3]{BruPedra-MFII}.

More generally, for possibly time-dependent interactions, the (generally
non-autonomous) local dynamics is defined, for any continuous function $\Psi
\in C(\mathbb{R};\mathcal{W}^{\mathbb{R}})$ and $L\in \mathbb{N}_{0}$, as
the unique (fundamental) solution $(\tau _{t,s}^{(L,\Psi )})_{_{s,t\in 
\mathbb{R}}}$ in the Banach space $\mathcal{B}(\mathcal{U})$ to the (finite
volume) non-auto%
\-%
nomous evolution equation\footnote{%
Let $\mathcal{H}$ be some Hilbert space and $(H_{t})_{t\in {\mathbb{R}}}$
some continuous family of bounded Hamiltonians acting on $\mathcal{H}$. The
corresponding Schr\"{o}dinger equation with $\hbar =1$ reads $i\partial
_{t}\varphi _{t}=H_{t}\varphi _{t}$ and so, $\varphi _{t}=\mathrm{U}%
_{t,s}\varphi _{s}$ with $\mathrm{U}_{t,s}$ being the solution to $\partial
_{t}\mathrm{U}_{t,s}=H_{t}\mathrm{U}_{t,s}$. Then, in the Heisenberg
picture, the time-evolution of any (bounded) observable $B$ acting on $%
\mathcal{H}$ at initial time $t=s\in {\mathbb{R}}$ is $B_{t}=\tau
_{t,s}\left( B_{s}\right) \doteq \mathrm{U}_{t,s}^{\ast }B_{s}\mathrm{U}%
_{t,s}$ for $s,t\in {\mathbb{R}}$, which yields $\partial _{t}\tau
_{t,s}=\tau _{t,s}\circ \delta _{t}$ with $\delta _{t}\left( A\right) \doteq
i[H_{t},A]$.}%
\begin{equation}
\forall s,t\in {\mathbb{R}}:\qquad \partial _{t}\tau _{t,s}^{(L,\Psi )}=\tau
_{t,s}^{(L,\Psi )}\circ \delta _{L}^{\Psi \left( t\right) }  \label{cauchy2}
\end{equation}%
with $\tau _{s,s}^{(L,\Psi )}=\mathbf{1}_{\mathcal{U}}$. The solution to (%
\ref{cauchy2}) can be explicitly written as a Dyson-Phillips series: For any 
$s,t\in {\mathbb{R}}$,%
\begin{eqnarray}
\tau _{t,s}^{(L,\Psi )} &=&\mathbf{1}_{\mathcal{U}}+\sum\limits_{k\in {%
\mathbb{N}}}\int_{s}^{t}\mathrm{d}t_{1}\cdots \int_{s}^{t_{k-1}}\mathrm{d}%
t_{k}  \label{finite dyson} \\
&&\qquad \qquad \qquad \delta _{L}^{\Psi \left( t_{k}\right) }\circ \cdots
\circ \delta _{L}^{\Psi \left( t_{1}\right) }.  \notag
\end{eqnarray}%
By \cite[Corollary 5.2]{brupedraLR}, in the thermodynamic limit $%
L\rightarrow \infty $, for any $\Psi \in C(\mathbb{R};\mathcal{W}^{\mathbb{R}%
})$, the group $(\tau _{t,s}^{(L,\Psi )})_{s,t\in {\mathbb{R}}}$, $L\in 
\mathbb{N}_{0}$, strongly converges, at any fixed $s,t$, to a strongly
continuous two-para%
\-%
meter family $(\tau _{t,s}^{\Psi })_{s,t\in {\mathbb{R}}}$ of $\ast $-auto%
\-%
morphisms of $\mathcal{U}$:%
\begin{equation}
\lim_{L\rightarrow \infty }\tau _{t,s}^{(L,\Psi )}(A)=:\tau _{t,s}^{\Psi
}\left( A\right) ,\qquad A\in \mathcal{U},\ s,t\in {\mathbb{R}}.
\label{dynamics limit}
\end{equation}%
In other words, (time-dependent) self-adjoint interactions lead to an
infinite volume (possibly non-autonomous) dynamics on the CAR algebra of the
lattice.

\subsubsection{Mean-Field Models\label{Long-Range Models}}

We start with some preliminary definitions: Let $\mathbb{S}$ be the unit
sphere of $\mathcal{W}_{1}$. For any $n\in \mathbb{N}$ and signed Borel
measure of finite variation $\mathfrak{a}$ on the Cartesian product $\mathbb{%
S}^{n}$ (endowed with its product topology), we define the signed Borel
measure (of finite variation) $\mathfrak{a}^{\ast }$ to be the pushforward
of $\mathfrak{a}$ through the self-homeomorphism 
\begin{equation}
\left( \Psi ^{(1)},\ldots ,\Psi ^{(n)}\right) \mapsto ((\Psi ^{(n)})^{\ast
},\ldots ,(\Psi ^{(1)})^{\ast })\in \mathbb{S}^{n}
\label{push forward self-adjoint}
\end{equation}%
of $\mathbb{S}^{n}$. A finite signed Borel measure $\mathfrak{a}$ on $%
\mathbb{S}^{n}$ is, by definition, \emph{self-adjoint} whenever $\mathfrak{%
a^{\ast }}=\mathfrak{a}$. For any $n\in \mathbb{N}$, the real Banach space
of self-adjoint signed Borel measures $\mathfrak{a}$ of finite variation on $%
\mathbb{S}^{n}$ with the norm of total variation%
\begin{equation*}
\Vert \mathfrak{a}\Vert _{\mathcal{S}(\mathbb{S}^{n}\mathbb{)}}:=|\mathfrak{a%
}|(\mathbb{S}^{n}),\qquad n\in \mathbb{N},
\end{equation*}%
is denoted by $\mathcal{S}(\mathbb{S}^{n}\mathbb{)}$. We define a norm for
sequences $\mathfrak{a}\equiv (\mathfrak{a}_{n})_{n\in \mathbb{N}}$ of
finite signed Borel measures $\mathfrak{a}_{n}\in \mathcal{S}(\mathbb{S}^{n}%
\mathbb{)}$ as follows: 
\begin{equation}
\left\Vert \mathfrak{a}\right\Vert _{\mathcal{S}}:=\sum_{n\in \mathbb{N}%
}n^{2}\mathbf{C}^{n-1}\left\Vert \mathfrak{a}_{n}\right\Vert _{\mathcal{S}(%
\mathbb{S}^{n}\mathbb{)}},\ \mathfrak{a}\equiv (\mathfrak{a}_{n})_{n\in 
\mathbb{N}}\in \mathcal{S},  \label{definition 0bis}
\end{equation}%
where the constant $\mathbf{C}>0$ is defined by (\ref{constant}). The
sequences which are finite in this norm form a (real) Banach space that we
denote by $\mathcal{S}$.

The (separable) Banach space of mean-field models is then defined by%
\begin{equation}
\mathcal{M}:=\mathcal{W}^{\mathbb{R}}\times \mathcal{S}
\label{def long range1}
\end{equation}%
with the norm of $\mathcal{M}$ being defined from (\ref{iteration00}) and (%
\ref{definition 0bis}) by%
\begin{equation}
\left\Vert \mathfrak{m}\right\Vert _{\mathcal{M}}:=\left\Vert \Phi
\right\Vert _{\mathcal{W}}+\left\Vert \mathfrak{a}\right\Vert _{\mathcal{S}%
},\quad \mathfrak{m}:=\left( \Phi ,\mathfrak{a}\right) \in \mathcal{M}.
\label{def long range2}
\end{equation}%
The spaces $\mathcal{W}^{\mathbb{R}}$ and $\mathcal{S}$ are seen as
subspaces of $\mathcal{M}$. In particular, $\Phi \equiv \left( \Phi
,0\right) \in \mathcal{M}$ for $\Phi \in \mathcal{W}^{\mathbb{R}}$. Using
the subspace $\mathcal{W}^{\mathrm{R}}$ of finite-range interactions defined
by (\ref{finite range}) for $\mathrm{R}\in \left[ 0,\infty \right) $, we
introduce the subspace 
\begin{eqnarray}
&&\mathcal{S}^{\infty }%
:=%
\bigcup\limits_{\mathrm{R}\in \left[ 0,\infty \right) }\left\{ (\mathfrak{a}%
_{n})_{n\in \mathbb{N}}\in \mathcal{S}:\right.   \label{S0bis} \\
&&\left. \forall n\in \mathbb{N},\ |\mathfrak{a}_{n}|(\mathbb{S}^{n})=|%
\mathfrak{a}_{n}|((\mathbb{S}\cap \mathcal{W}^{\mathrm{R}})^{n})\right\} . 
\notag
\end{eqnarray}%
Long-range dynamics are studied for models in the following two subspaces 
\begin{equation}
\mathcal{M}^{\infty }:=\mathcal{W}^{\mathbb{R}}\times \mathcal{S}^{\infty
},\ \mathcal{M}_{1}^{\infty }:=\left( \mathcal{W}^{\mathbb{R}}\cap \mathcal{W%
}_{1}\right) \times \mathcal{S}^{\infty }.  \label{M infinit}
\end{equation}%
Clearly, $\mathcal{W}^{\mathbb{R}}\subseteq \mathcal{M}^{\infty }\subseteq 
\mathcal{M}$ and $\left( \mathcal{W}^{\mathbb{R}}\cap \mathcal{W}_{1}\right)
\subseteq \mathcal{M}_{1}^{\infty }\subseteq \mathcal{M}^{\infty }$.

Similar to self-adjoint short-range interactions, each mean-field model
leads to a sequence of local Hamiltonians:\ For any $L\in \mathbb{N}_{0}$
and $\mathfrak{m}\in \mathcal{M}$,%
\begin{eqnarray}
&&U_{L}^{\mathfrak{m}}%
:=%
U_{L}^{\Phi }+\sum_{n\in \mathbb{N}}\frac{1}{\left\vert \Lambda
_{L}\right\vert ^{n-1}}  \label{equation long range energy} \\
&&\int_{\mathbb{S}^{n}}U_{L}^{\Psi ^{(1)}}\cdots U_{L}^{\Psi ^{(n)}}%
\mathfrak{a}_{n}\left( \mathrm{d}\Psi ^{(1)},\ldots ,\mathrm{d}\Psi
^{(n)}\right)  \notag
\end{eqnarray}%
with $U_{L}^{\Phi }$ and $U_{L}^{\Psi ^{(k)}}$ been defined by (\ref{local
energy}). Note that $U_{L}^{\mathfrak{m}}=(U_{L}^{\mathfrak{m}})^{\ast }$
and straightforward estimates yield the upper bound 
\begin{equation}
\left\Vert U_{L}^{\mathfrak{m}}\right\Vert _{\mathcal{U}}\leq \mathbf{C}%
\left\vert \Lambda _{L}\right\vert \left\Vert \mathfrak{m}\right\Vert _{%
\mathcal{M}},\qquad L\in \mathbb{N}_{0}.  \label{energy bound long range}
\end{equation}%
Compare with (\ref{norm Uphi}).

Similar to (\ref{quantum dyn local}), each model $\mathfrak{m}\in \mathcal{M}
$ leads to finite volume dynamics defined, for any $L\in \mathbb{N}_{0}$, by 
\begin{equation}
\tau _{t}^{(L,\mathfrak{m})}(A)=\mathrm{e}^{itU_{L}^{\mathfrak{m}}}A\mathrm{e%
}^{-itU_{L}^{\mathfrak{m}}},\qquad A\in \mathcal{U}.
\label{finite vol long range dyna}
\end{equation}%
In contrast with short range interactions (see (\ref{dynamics limit})), for
any fixed $A\in \mathcal{U}$ and $t\in \mathbb{R}$, the thermodynamic limit $%
L\rightarrow \infty $ of $\tau _{t}^{(L,\mathfrak{m})}(A)$ does \emph{not}
necessarily exist in the $C^{\ast }$-algebra $\mathcal{U}$. However, by \cite%
[Theorem 4.3]{BruPedra-MFIII}, for any $\mathfrak{m}\in \mathcal{M}%
_{1}^{\infty }$, it converges in the $\sigma $-weak topology within some
representation of $\mathcal{U}$. This is reminiscent of the fact that\ the
energy-density observable $U_{L}^{\Phi }/\left\vert \Lambda _{L}\right\vert $
does not converges in $\mathcal{U}$, as $L\rightarrow \infty $, but its
expectation value with respect to any periodic state does. See Appendix \ref%
{Long-Range Dynamics} for more details.

\subsection{State Spaces}

\subsubsection{Finite Volume State Space\label{Finite-Volume State Spaces}}

For any (non-empty) finite subset $\Lambda \subseteq \mathbb{Z}^{d}$, i.e., $%
\Lambda \in \mathcal{P}_{f}$, let $\mathcal{U}_{\Lambda }^{\ast }$ be the
dual space of the local $C^{\ast }$-algebra $\mathcal{U}_{\Lambda }$. For
every $\Lambda \in \mathcal{P}_{f}$, we denote by%
\begin{equation}
E_{\Lambda }:=\{\rho _{\Lambda }\in \mathcal{U}_{\Lambda }^{\ast }:\rho
_{\Lambda }\geq 0,\ \rho _{\Lambda }(\mathfrak{1})=1\}  \label{local states}
\end{equation}%
the space of all states on $\mathcal{U}_{\Lambda }$. For all $\Lambda \in 
\mathcal{P}_{f}$, the space $E_{\Lambda }$ is a norm-compact convex subset
of the dual space $\mathcal{U}_{\Lambda }^{\ast }$ and, for any $\rho
_{\Lambda }\in E_{\Lambda }$, there is a unique, positive, trace-one
operator $\mathrm{d}_{\Lambda }\in \mathcal{B}(\mathcal{F}_{\Lambda _{L}})$
satisfying 
\begin{equation}
\rho _{\Lambda }\left( A\right) :=\mathrm{Trace}_{\mathcal{F}_{\Lambda
_{L}}}\left( \mathrm{d}_{\Lambda }A\right) ,\qquad A\in \mathcal{U}_{\Lambda
},  \label{density matrix}
\end{equation}%
named the density matrix of $\rho _{\Lambda }$. In fact, $E_{\Lambda }$ is
affinely equivalent to the set of all states on the algebra of $%
2^{\left\vert \Lambda \right\vert \times \left\vert \mathrm{S}\right\vert
}\times 2^{\left\vert \Lambda \right\vert \times \left\vert \mathrm{S}%
\right\vert }$ complex matrices. The structure of states for infinite
systems is more subtle, as demonstrated in \cite%
{Bru-pedra-MF-I,Bru-pedra-hypertopo}.

Note that the physically relevant finite volume states $\rho _{\Lambda }$, $%
\Lambda \in \mathcal{P}_{f}$, are \emph{even}, i.e., $\rho _{\Lambda }\circ
\sigma |_{\mathcal{U}_{\Lambda }}=\rho _{\Lambda }$ with $\sigma |_{\mathcal{%
U}_{\Lambda }}$ being the restriction to $\mathcal{U}_{\Lambda }$ of the
unique $\ast $-automorphism $\sigma $ of $\mathcal{U}$ satisfying (\ref%
{automorphism gauge invariance}). It means that $\rho _{\Lambda }$ vanishes
on all \emph{odd} monomials in $\{a_{x,\mathrm{s}},a_{x,\mathrm{s}}^{\ast
}\}_{x\in \Lambda ,\mathrm{s}\in \mathrm{S}}$. For any $\Lambda \in \mathcal{%
P}_{f}$, we define 
\begin{equation}
E_{\Lambda }^{+}:=\left\{ \rho _{\Lambda }\in E_{\Lambda }:\rho _{\Lambda
}\circ \sigma |_{\mathcal{U}_{\Lambda }}=\rho _{\Lambda }\right\} \subseteq
E_{\Lambda }  \label{finite volume even states}
\end{equation}%
as being the space of all finite volume even states.

\subsubsection{Infinite Volume State Spaces}

For the infinite system, let $\mathcal{U}^{\ast }\equiv \mathcal{U}_{\mathbb{%
Z}^{d}}^{\ast }$ be the dual space of $\mathcal{U}\equiv \mathcal{U}_{%
\mathbb{Z}^{d}}$. In contrast with $\mathcal{U}_{\Lambda }$, $\Lambda \in 
\mathcal{P}_{f}$, $\mathcal{U}$ has infinite dimension and the natural
topology on $\mathcal{U}^{\ast }$ is the weak$^{\ast }$\ topology\footnote{%
Recal that the weak$^{\ast }$ topology of $\mathcal{U}^{\ast }$ is the
coarsest topology on $\mathcal{U}^{\ast }$ that makes the mapping $\rho
\mapsto \rho \left( A\right) $ continuous for every $A\in \mathcal{U}$. See 
\cite[Sections 3.8, 3.10, 3.14]{Rudin} for more details.} (and not the norm
topology). Thus, the topology of $\mathcal{U}^{\ast }$ considered here is
always the weak$^{\ast }$\ topology and, in this case, $\mathcal{U}^{\ast }$
is a locally convex space, by \cite[Theorem 3.10]{Rudin}.

Similar to (\ref{local states}), the state space on $\mathcal{U}$ is defined
by%
\begin{equation}
E\equiv E_{\mathbb{Z}^{d}}:=\{\rho \in \mathcal{U}^{\ast }:\rho \geq 0,\
\rho (\mathfrak{1})=1\}.  \label{states CAR}
\end{equation}%
It is a metrizable, convex and compact subset of $\mathcal{U}^{\ast }$, by 
\cite[Theorems 3.15 and 3.16]{Rudin}. It is also the state space of the
classical dynamics we define in \cite{Bru-pedra-MF-I}. By the Krein-Milman
theorem \cite[Theorem 3.23]{Rudin}, $E$ is the closure of the convex hull of
the (non-empty) set of its extreme points, which are meanwhile dense in $E$,
by \cite[Eq. (13)]{BruPedra-MFII}.

As explained below Equation (\ref{definition of even operators}), recall
that the $C^{\ast }$-algebra $\mathcal{U}^{+}$ should be considered as more
fundamental than $\mathcal{U}$ in Physics, because of the principle of
locality in quantum field theory. As a consequence, states on the $C^{\ast }$%
-algebra $\mathcal{U}^{+}$ should be seen as being the physically relevant
ones. The set of states on $\mathcal{U}^{+}$ is canonically identified with
the metrizable, convex and compact set of \emph{even} states defined by%
\begin{eqnarray}
E^{+} &\equiv &E_{\mathbb{Z}^{d}}^{+}%
:=%
\left\{ \rho \in \mathcal{U}^{\ast }:\rho \geq 0,\ \right.
\label{gauge invariant states} \\
&&\left. \qquad \qquad \rho (\mathfrak{1})=1,\ \rho \circ \sigma =\rho
\right\} ,  \notag
\end{eqnarray}%
$\sigma $ being the unique $\ast $-automorphism of $\mathcal{U}$ satisfying (%
\ref{automorphism gauge invariance}). This space and the full state space, $%
E $, are equivalent convex topological spaces, i.e., there is an affine
homeomorphism between them. In particular, $E^{+}$ is the closure of the
convex hull of the (non-empty) set of its extreme points, which are dense in 
$E^{+}$. See \cite[Proposition 2.1]{BruPedra-MFII} and its proof.

Note that the spaces $E$ and $E^{+}$, having a dense set of extreme points\
-- i.e., having a dense extreme boundary -- has a much more peculiar
structure than the finite volume state space $E_{\Lambda }$ for $\Lambda \in 
\mathcal{P}_{f}$. At first glance, it may look very strange for a non-expert
on convex analysis, but it should not be that surprising: For instance, the
unit ball of any infinite-dimensional Hilbert space has clearly a dense
extreme boundary in the weak topology. In fact, the existence of convex
compact sets with dense extreme boundary \emph{is not an accident} in
infinite-dimensional spaces, but rather generic, in the topological sense.
This has been first proven \cite{Klee} in 1959 for convex norm-compact sets
within a separable Banach space. Recently, in \cite[Section 2.3]%
{Bru-pedra-MF-I} and more generally in \cite{Bru-pedra-hypertopo}, the
property of having dense extreme boundary is proven to be generic for weak$%
^{\ast }$-compact convex sets within the dual space of an
infinite-dimensional topological space. As a matter of fact, all state
spaces of infinite volume systems one is going to encounter in the current
paper have a dense extreme boundary, except the set of permutation-invariant
states described in Appendix \ref{sect Permutation-Invariant Long-Range
Models}, because these can be reduced to states of finite-dimensional matrix
algebras, via de Finetti-type results.

\subsubsection{Periodic State Spaces\label{section periodic states}}

Consider the subgroups $(\mathbb{Z}_{\vec{\ell}}^{d},+)\subseteq (\mathbb{Z}%
^{d},+)$, $\vec{\ell}\in \mathbb{N}^{d}$, where%
\begin{equation*}
\mathbb{Z}_{\vec{\ell}}^{d}:=\ell _{1}\mathbb{Z}\times \cdots \times \ell
_{d}\mathbb{Z}.
\end{equation*}%
At fixed $\vec{\ell}\in \mathbb{N}^{d}$, a state $\rho \in E$ satisfying $%
\rho \circ \alpha _{x}=\rho $ for all $x\in \mathbb{Z}_{\vec{\ell}}^{d}$ is
called $\mathbb{Z}_{\vec{\ell}}^{d}$\emph{-invariant} on $\mathcal{U}$ or $%
\vec{\ell}$\emph{-periodic}, $\alpha _{x}$ being the unique $\ast $%
-automorphism of $\mathcal{U}$ satisfying (\ref{transl}).
Translation-invariant states refer to $(1,\ldots ,1)$-periodic states. For
any $\vec{\ell}\in \mathbb{N}^{d}$, let 
\begin{equation}
E_{\vec{\ell}}:=\left\{ \rho \in E:\rho \circ \alpha _{x}=\rho \quad \text{%
for}\ \text{all}\ x\in \mathbb{Z}_{\vec{\ell}}^{d}\right\}
\label{periodic invariant states}
\end{equation}%
be the space of $\vec{\ell}$-periodic states. By \cite[Lemma 1.8]{BruPedra2}%
, periodic states are always even and, by \cite[Proposition 2.3]%
{BruPedra-MFII}, the set of all periodic states%
\begin{equation}
E_{\mathrm{p}}:=\bigcup_{\vec{\ell}\in \mathbb{N}^{d}}E_{\vec{\ell}%
}\subseteq E^{+}  \label{set of periodic states}
\end{equation}%
is dense in the space $E^{+}$ of even states.

For each $\vec{\ell}\in \mathbb{N}^{d}$, $E_{\vec{\ell}}$\ is metrizable,
convex and compact and, by the Krein-Milman theorem \cite[Theorem 3.23]%
{Rudin}, it is the closure of the convex hull of the (non-empty) set $%
\mathcal{E}_{\vec{\ell}}$ of its extreme points, which is a $G_{\delta }$
subset of $E_{\vec{\ell}}$ (in particular it is Borel measurable). In fact,
by \cite[Theorem 1.9]{BruPedra2} (which uses the Choquet theorem \cite[p. 14]%
{Phe}), for any $\rho \in E_{\vec{\ell}}$, there is a unique probability
measure $\mu _{\rho }$ on $E_{\vec{\ell}}$ with support in $\mathcal{E}_{%
\vec{\ell}}$ such that\footnote{%
The integral in (\ref{affine decomposition0}) means that $\rho \in E_{\vec{%
\ell}}$ is the (unique) barycenter of the normalized positive Borel regular
measure $\mu _{\rho }$ on $E_{\vec{\ell}}$. See, e.g., \cite[Definition
10.15, Theorem 10.16, and also Lemma 10.17]{BruPedra2}.}%
\begin{equation}
\rho =\int_{\mathcal{E}_{\vec{\ell}}}\hat{\rho}\ \mathrm{d}\mu _{\rho
}\left( \hat{\rho}\right) .  \label{affine decomposition0}
\end{equation}%
The set $\mathcal{E}_{\vec{\ell}}$ can be characterized by an ergodicity
property of states, see \cite[Theorem 1.16]{BruPedra2}. Moreover, $\mathcal{E%
}_{\vec{\ell}}$ is dense in $E_{\vec{\ell}}$, by \cite[Corollary 4.6]%
{BruPedra2}. In other words, like the sets $E$ and $E^{+}$, $E_{\vec{\ell}}$
has dense extreme boundary for any $\vec{\ell}\in \mathbb{N}^{d}$.

\subsection{Long-Range Dynamics\label{Long-Range Dynamics}}

\subsubsection{Self-Consistency Equations}

Generically, as already discussed in the main part of the paper, mean-field
dynamics in infinite volume are intricate combinations of a classical and
quantum dynamics. Similar to \cite[Theorem 4.1]{Bru-pedra-MF-I}, both
dynamics are related the existence of a solution to a (dynamical) \emph{%
self-consistency equation}. In order to present such equations we need some
preliminary definitions: For $\vec{\ell}\in \mathbb{N}^{d}$, $\mathfrak{m}%
=(\Phi ,\mathfrak{\mathfrak{a}})\in \mathcal{M}$ and $\rho \in E$, we define
the approximating (self-adjoint, short-range) interaction $\Phi ^{(\mathfrak{%
m},\rho )}\in \mathcal{W}^{\mathbb{R}}$ by 
\begin{eqnarray}
&&\Phi ^{(\mathfrak{m},\rho )}%
:=%
\Phi +\sum_{n\in \mathbb{N}}\int_{\mathbb{S}^{n}}\mathfrak{a}_{n}\left( 
\mathrm{d}\Psi ^{(1)},\ldots ,\mathrm{d}\Psi ^{(n)}\right)  \notag \\
&&\sum_{m=1}^{n}\Psi ^{(m)}\prod\limits_{j\in \left\{ 1,\ldots ,n\right\}
,j\neq m}\rho (\mathfrak{e}_{\Psi ^{(j)},\vec{\ell}}),
\label{approx interaction}
\end{eqnarray}%
where 
\begin{eqnarray}
&&\mathfrak{e}_{\Phi ,\vec{\ell}}%
:=%
\frac{1}{\ell _{1}\cdots \ell _{d}}\sum\limits_{x=(x_{1},\ldots
,x_{d}),\;x_{i}\in \{0,\ldots ,\ell _{i}-1\}}  \notag \\
&&\qquad \qquad \qquad \sum\limits_{\Lambda \in \mathcal{P}_{f},\;\Lambda
\ni x}\frac{\Phi _{\Lambda }}{\left\vert \Lambda \right\vert }.
\label{eq:enpersite}
\end{eqnarray}%
Recall meanwhile that $\mathcal{M}^{\infty }:=\mathcal{W}^{\mathbb{R}}\times 
\mathcal{S}^{\infty }$, see (\ref{S0bis})-(\ref{M infinit}). Then, by \cite[%
Theorem 6.5]{BruPedra-MFII}, for any $\mathfrak{m}\in \mathcal{M}^{\infty }$%
, there is a unique continuous\footnote{%
We endow the set $C\left( E;E\right) $ of continuous functions from $E$ to
itself with the topology of uniform convergence. See \cite[Eq. (100)]%
{BruPedra-MFII} for more details.} mapping $\mathbf{\varpi }^{\mathfrak{m}}$
from $\mathbb{R}$ to the space of automorphisms\footnote{%
I.e., elements of $C\left( E;E\right) $ with continuous inverse.} (or
self-homeomorphisms) of $E$ such that 
\begin{equation}
\mathbf{\varpi }^{\mathfrak{m}}\left( t;\rho \right) =\rho \circ \tau
_{t,0}^{\Psi ^{\left( \mathfrak{m},\rho \right) }},\qquad t\in {\mathbb{R}}%
,\ \rho \in E,  \label{self-consistency equation}
\end{equation}%
with $\Psi ^{\left( \mathfrak{m},\rho \right) }\in C(\mathbb{R};\mathcal{W}^{%
\mathbb{R}})$, $\rho \in E$, defined by 
\begin{equation}
\Psi ^{\left( \mathfrak{m},\rho \right) }(t):=\Phi ^{(\mathfrak{m},\mathbf{%
\varpi }^{\mathfrak{m}}\left( t;\rho \right) )},\qquad t\in {\mathbb{R}},
\label{self-consistency equation2}
\end{equation}%
and where the strongly continuous two-para%
\-%
meter family $(\tau _{t,s}^{\Psi ^{\left( \mathfrak{m},\rho \right)
}})_{s,t\in {\mathbb{R}}}$ is the strong limit, at any fixed $s,t\in {%
\mathbb{R}}$, of the local dynamics $(\tau _{t,s}^{(L,\Psi ^{\left( 
\mathfrak{m},\rho \right) })})_{s,t\in {\mathbb{R}}}$ defined by (\ref%
{cauchy2}) for $\Psi =\Psi ^{\left( \mathfrak{m},\rho \right) }$, see (\ref%
{dynamics limit}) and \cite[Corollary 5.2]{brupedraLR}. Equation (\ref%
{self-consistency equation}) is named here the (dynamical) self-consistency
equation.

\subsubsection{Quantum Part of Mean-Field Dynamics\label{Quantum Part of
Long-Range Dynamics}}

Recall that any model $\mathfrak{m}\in \mathcal{M}$ leads to finite volume
dynamics $(\tau _{t}^{(L,\mathfrak{m})})_{t\in \mathbb{R}}$, $L\in \mathbb{N}%
_{0}$, defined by (\ref{finite vol long range dyna}). Therefore, at $L\in 
\mathbb{N}_{0}$, the time-evolution $(\rho _{t}^{(L)})_{t\in \mathbb{R}}$ of
any finite volume state $\rho ^{(L)}\in E_{\Lambda _{L}}$ is given by 
\begin{equation}
\rho _{t}^{(L)}:=\rho ^{(L)}\circ \tau _{t}^{(L,\mathfrak{m})}.
\label{long-range dyn0}
\end{equation}%
The corresponding time-dependent density matrix is $\mathrm{d}%
_{t}^{(L)}=\tau _{-t}^{(L,\mathfrak{m})}(\mathrm{d}^{(L)})$. Equation (\ref%
{long-range dyn0}) refers to the Schr\"{o}dinger picture of quantum
mechanics.

As already mentioned, for any fixed $A\in \mathcal{U}$ and $t\in \mathbb{R}$%
, the thermodynamic limit $L\rightarrow \infty $ of $\tau _{t}^{(L,\mathfrak{%
m})}(A)$ does not necessarily exist in $\mathcal{U}$, but the limit $%
L\rightarrow \infty $ of $\rho _{t}^{(L)}$ can still make sense: Fix once
and for all a translation-invariant model $\mathfrak{m}\in \mathcal{M}%
_{1}^{\infty }$, see (\ref{M infinit}). Take $\vec{\ell}\in \mathbb{N}^{d}$
and recall that $E_{\vec{\ell}}$ is the space of $\vec{\ell}$-periodic
states defined by (\ref{periodic invariant states}), whose set of extreme
points is denoted $\mathcal{E}_{\vec{\ell}}$. Recall also (\ref{affine
decomposition0}), i.e., that, for any $\rho \in E_{\vec{\ell}}$, there is a
unique probability measure $\mu _{\rho }$ on $E_{\vec{\ell}}$ with support
in $\mathcal{E}_{\vec{\ell}}$ such that%
\begin{equation*}
\rho =\int_{\mathcal{E}_{\vec{\ell}}}\hat{\rho}\ \mathrm{d}\mu _{\rho
}\left( \hat{\rho}\right) .
\end{equation*}%
From the fact that the set $\mathcal{E}_{\vec{\ell}}$ is characterized by an
ergodicity property (see \cite[Theorem 1.16]{BruPedra2}), one can prove
that, for any $A\in \mathcal{U}$, 
\begin{eqnarray}
&&\lim_{L\rightarrow \infty }\rho \circ \tau _{t}^{(L,\mathfrak{m})}\left(
A\right)  \notag \\
&=&\int_{\mathcal{E}_{\vec{\ell}}}\mathbf{\varpi }^{\mathfrak{m}}\left( t;%
\hat{\rho}\right) \left( A\right) \ \mathrm{d}\mu _{\rho }\left( \hat{\rho}%
\right)  \notag \\
&=&\int_{\mathcal{E}_{\vec{\ell}}}\hat{\rho}\circ \tau _{t,0}^{\Psi ^{\left( 
\mathfrak{m},\hat{\rho}\right) }}\left( A\right) \ \mathrm{d}\mu _{\rho
}\left( \hat{\rho}\right) ,  \label{long-range dyn}
\end{eqnarray}%
where $\mathbf{\varpi }^{\mathfrak{m}}$ is the solution to the
self-consistency equation\emph{\ }(\ref{self-consistency equation}). See 
\cite[Proposition 4.2, Theorem 4.3]{BruPedra-MFIII}. Using in particular,
for any $L\in \mathbb{N}_{0}$, the restriction $\rho ^{(L)}:=\rho |_{%
\mathcal{U}_{\Lambda _{L}}}$ of a state $\rho \in E_{\vec{\ell}}$ to $%
\mathcal{U}_{\Lambda _{L}}$ then (\ref{long-range dyn}) can also be seen as
the thermodynamic limit $L\rightarrow \infty $ of the expectation value $%
\rho _{t}^{(L)}(A)$ of any local element $A\in \mathcal{U}_{0}$, the
time-dependent state $\rho _{t}^{(L)}$ been defined by (\ref{long-range dyn0}%
).

Equation (\ref{long-range dyn}) means in fact that the thermodynamic limit $%
L\rightarrow \infty $ of $\tau _{t}^{(L,\mathfrak{m})}(A)$ exists in the
GNS\ representation\footnote{%
Recall that $\mathcal{H}_{\rho }$ is an\ Hilbert space, $\pi _{\rho }:%
\mathcal{U}\rightarrow \mathcal{B}\left( \mathcal{H}_{\rho }\right) $ is a
representation of $\mathcal{U}$ and $\Omega _{\rho }\in \mathcal{H}_{\rho }$
is a cyclic vector for $\pi _{\rho }\left( \mathcal{U}\right) $.} $\left( 
\mathcal{H}_{\rho },\pi _{\rho },\Omega _{\rho }\right) $ of $\mathcal{U}$
associated with the initial state $\rho $. More precisely, one obtains a
dynamics $(T_{t}^{\mathfrak{m}})_{t\in {\mathbb{R}}}$ defined by%
\begin{equation*}
T_{t}^{\mathfrak{m}}\circ \pi _{\rho }\left( A\right) =\lim_{L\rightarrow
\infty }\pi _{\rho }\circ \tau _{t}^{(L,\mathfrak{m})}\left( A\right)
,\qquad A\in \mathcal{U},
\end{equation*}%
on the (von Neumann) subalgebra $\pi _{\rho }(\mathcal{U)}^{\prime \prime }$
of the algebra $\mathcal{B}\left( \mathcal{H}_{\rho }\right) $ of bounded
operators on the Hilbert space $\mathcal{H}_{\rho }$. The above limit has to
be understood in the $\sigma $-weak topology within $\mathcal{B}(\mathcal{H}%
_{\rho })$ (and in many cases one could even prove strong convergence). This
refers to the quantum part of the mean-field dynamics (in some
representation), which is generally \emph{non-autonomous}, although the
primordial local dynamics is autonomous.

\subsubsection{Classical Part of Mean-Field Dynamics\label{Section classical
dynamics-general}}

For any $\vec{\ell}\in \mathbb{N}^{d}$, the infinite volume mean-field
dynamics of $\vec{\ell}$-periodic states, as given by (\ref{long-range dyn}%
), involves the knowledge of a continuous flow\footnote{%
That is, the continuous mapping $\mathbf{\varpi }^{\mathfrak{m}}$ from $%
\mathbb{R}$ to the space of automorphisms (or self-homeomorphisms) of $E$
defined by (\ref{self-consistency equation}).} on $\mathcal{E}_{\vec{\ell}}$%
. Seeing $\mathcal{E}_{\vec{\ell}}$ or $E_{\vec{\ell}}=\overline{\mathcal{E}%
_{\vec{\ell}}}$ as a (classical) phase space, it becomes natural to study
the classical Hamiltonian dynamics associated with this flow, as is usual in
classical mechanics. Note that, for a (possibly non-translation-invariant)
model $\mathfrak{m}\in \mathcal{M}^{\infty }$, for any $t\in \mathbb{R}$, $%
\mathbf{\varpi }^{\mathfrak{m}}\left( t;\cdot \right) $ preserves the space $%
E^{+}$ of \emph{even} states defined by (\ref{gauge invariant states}), but
not necessarily $E_{\vec{\ell}}$. If $\mathfrak{m}\in \mathcal{M}%
_{1}^{\infty }$ then, for any $\vec{\ell}\in \mathbb{N}^{d}$, the flow lets
the sets $\mathcal{E}_{\vec{\ell}}$ and $E_{\vec{\ell}}$ invariant. See (\ref%
{conservation}) below. Here, we adopt a broader perspective by taking the
full state space $E$, defined by (\ref{states CAR}), because the classical
dynamics described below can be easily pushed forward, through the
restriction map, from $C(E;\mathbb{C})$ to $C(E^{+};\mathbb{C})$ for general 
$\mathfrak{m}\in \mathcal{M}^{\infty }$, and also to $C(E_{\vec{\ell}};%
\mathbb{C})$ for any $\vec{\ell}\in \mathbb{N}^{d}$, when $\mathfrak{m}\in 
\mathcal{M}_{1}^{\infty }$ is translation-invariant.

Note that $C(E;\mathbb{C})$, $C(E^{+};\mathbb{C})$ and $C(E_{\vec{\ell}};%
\mathbb{C})$, endowed with the point-wise operations and complex conjugation
as well as the supremum norm, are unital commutative $C^{\ast }$-algebras.
For any model $\mathfrak{m}\in \mathcal{M}^{\infty }$, the mapping $\mathbf{%
\varpi }^{\mathfrak{m}}$, the solution to the self-consistency equation\emph{%
\ }(\ref{self-consistency equation}), yields a family $(V_{t}^{\mathfrak{m}%
})_{t\in \mathbb{R}}$ of $\ast $-automorphisms on $C(E;\mathbb{C})$ defined
by 
\begin{equation}
V_{t}^{\mathfrak{m}}\left( f\right) :=f\circ \mathbf{\varpi }^{\mathfrak{m}%
}\left( t\right) ,\quad f\in C\left( E;\mathbb{C}\right) ,\ t\in \mathbb{R}.
\label{classical evolution family}
\end{equation}%
It is a Feller group: $(V_{t}^{\mathfrak{m}})_{t\in \mathbb{R}}$ is a
strongly continuous group of $\ast $-automorphisms of $C(E;\mathbb{C})$,
which is obviously positivity preserving and has operator norm equal to one.
See \cite[Proposition 6.8]{BruPedra-MFII}. When restricted to the dense
subspace $E_{\mathrm{p}}\subseteq E^{+}$ (\ref{set of periodic states}) of
all periodic states, the ones we are interested in (cf. (\ref{long-range dyn}%
)), for any translation-invariant model $\mathfrak{m}\in \mathcal{M}%
_{1}^{\infty }$, the one-parameter group $(V_{t}^{\mathfrak{m}})_{t\in 
\mathbb{R}}$ is generated by a Poissonian symmetric derivation: \bigskip

\noindent \underline{(i) Local polynomials:} Elements of the$\ C^{\ast }$%
-algebra $\mathcal{U}$ naturally define continuous and affine functions $%
\hat{A}\in C(E;\mathbb{C})$ by 
\begin{equation*}
\hat{A}\left( \rho \right) :=\rho \left( A\right) ,\qquad \rho \in E,\ A\in 
\mathcal{U}.
\end{equation*}%
This is the well-known Gelfand transform. Recall that $\mathcal{U}_{0}$ is
the normed $\ast $-algebra of local elements of $\mathcal{U}$ defined by (%
\ref{simple}). We denote by%
\begin{equation}
\mathbb{P}:=\mathbb{C}[\{\hat{A}:A\in \mathcal{U}_{0}\}]\subseteq C(E;%
\mathbb{C})  \label{polynomials}
\end{equation}%
the subspace of (local) polynomials in the elements of $\{\hat{A}:A\in 
\mathcal{U}_{0}\}$, with complex coefficients. \bigskip

\noindent \underline{(ii) Poisson structure:} For any $n\in \mathbb{N}$, $%
A_{1},\ldots ,A_{n}\in \mathcal{U}$ and $g\in C^{1}\left( \mathbb{R}^{n},%
\mathbb{C}\right) $ we define the function $\Gamma _{g}\in C(E;\mathbb{C})$
by%
\begin{equation}
\Gamma _{g}\left( \rho \right) :=g\left( \rho \left( A_{1}\right) ,\ldots
,\rho \left( A_{n}\right) \right) ,\qquad \rho \in E.  \label{ddddd}
\end{equation}%
Functions of this type are known in the literature as cylindrical functions.
For such a function and any $\rho \in E$, define 
\begin{eqnarray}
&&\mathrm{D}\Gamma _{g}\left( \rho \right) 
:=%
\sum_{j=1}^{n}\left( A_{j}-\rho \left( A_{j}\right) \mathfrak{1}\right)
\label{Ynbis} \\
&&\qquad \qquad \partial _{x_{j}}g\left( \rho \left( A_{1}\right) ,\ldots
,\rho \left( A_{n}\right) \right)  \notag
\end{eqnarray}%
for any $\rho \in E$. This definition comes from a notion, introduced by us,
of a convex weak$^{\ast }$-continuous Gateaux derivative, as explained in 
\cite[Section 5.2]{BruPedra-MFII}. Then, for any $n,m\in \mathbb{N}$, $%
A_{1},\ldots ,A_{n},B_{1},\ldots ,B_{m}\in \mathcal{U}$, $g\in C^{1}\left( 
\mathbb{R}^{n},\mathbb{C}\right) $ and $h\in C^{1}\left( \mathbb{R}^{m},%
\mathbb{C}\right) $, we define the continuous function $\left\{ \Gamma
_{h},\Gamma _{g}\right\} \in C(E;\mathbb{C})$ by%
\begin{equation}
\left\{ \Gamma _{h},\Gamma _{g}\right\} \left( \rho \right) :=\rho \left( i 
\left[ \mathrm{D}\Gamma _{h}\left( \rho \right) ,\mathrm{D}\Gamma _{g}\left(
\rho \right) \right] \right)  \label{poisson bracket}
\end{equation}%
for any $\rho \in E$, where $A_{1},\ldots ,A_{n}\in \mathcal{U}$ and $%
B_{1},\ldots ,B_{m}$ respectively determine $\Gamma _{g}$ and $\Gamma _{h}$
via (\ref{ddddd}). This defines a Poisson bracket on the space $\mathbb{P}$
of all (local) polynomial functions\ acting on $E$. By construction, for any 
$\vec{\ell}\in \mathbb{N}^{d}$,%
\begin{equation}
\begin{array}{l}
\left\{ \Gamma _{h}|_{E^{+}},\Gamma _{g}|_{E^{+}}\right\} :=\left\{ \Gamma
_{h},\Gamma _{g}\right\} |_{E^{+}} \\ 
\{\Gamma _{h}|_{E_{\vec{\ell}}},\Gamma _{g}|_{E_{\vec{\ell}}}\}:=\left\{
\Gamma _{h},\Gamma _{g}\right\} |_{E_{\vec{\ell}}} \\ 
\{\Gamma _{h}|_{\mathcal{E}_{\vec{\ell}}},\Gamma _{g}|_{\mathcal{E}_{\vec{%
\ell}}}\}:=\left\{ \Gamma _{h},\Gamma _{g}\right\} |_{\mathcal{E}_{\vec{\ell}%
}}%
\end{array}
\label{poisson bracketsbis}
\end{equation}%
also define a Poisson bracket on polynomials of $C(E^{+};\mathbb{C})$, $C(E_{%
\vec{\ell}};\mathbb{C})$ and $C(\mathcal{E}_{\vec{\ell}};\mathbb{C})$,
respectively. This definition can be extended to the space 
\begin{equation*}
\mathfrak{Y}\equiv C^{1}\left( E;\mathbb{C}\right) \subseteq C\left( E;%
\mathbb{C}\right)
\end{equation*}%
of continuously differentiable functions. See \cite[Section 5.2]%
{BruPedra-MFII} and \cite[Section 3]{Bru-pedra-MF-I} for a more detailed
construction of such Poisson structures. \bigskip

\noindent \underline{(iii) Liouville's equation:} Local classical energy
functions \cite[Definition 6.9]{BruPedra-MFII} associated with $\mathfrak{m}%
\in \mathcal{M}$ are defined, for any $L\in \mathbb{N}_{0}$, by%
\begin{eqnarray}
&&\mathrm{h}_{L}^{\mathfrak{m}}%
:=%
\widehat{U_{L}^{\Phi }}+\sum_{n\in \mathbb{N}}\frac{1}{\left\vert \Lambda
_{L}\right\vert ^{n-1}}\int_{\mathbb{S}^{n}}  \label{classical energy} \\
&&\widehat{U_{L}^{\Psi ^{(1)}}}\cdots \widehat{U_{L}^{\Psi ^{(n)}}}\mathfrak{%
a}_{n}\left( \mathrm{d}\Psi ^{(1)},\ldots ,\mathrm{d}\Psi ^{(n)}\right) . 
\notag
\end{eqnarray}%
Note that $\mathrm{h}_{L}^{\mathfrak{m}}\in C^{1}\left( E;\mathbb{C}\right) $%
. Compare with the local Hamiltonian $U_{L}^{\mathfrak{m}}$ defined by (\ref%
{equation long range energy}). Then, by \cite[Corollary 6.12]{BruPedra-MFII}%
, for each translation-invariant model $\mathfrak{m}\in \mathcal{M}%
_{1}^{\infty }$, any time $t\in \mathbb{R}$ and all local polynomials $f\in 
\mathbb{P}$, one has $V_{t}^{\mathfrak{m}}(f)\in C^{1}(E;\mathbb{C})$ and%
\begin{eqnarray}
\partial _{t}V_{t}^{\mathfrak{m}}\left( f\right) &=&V_{t}^{\mathfrak{m}%
}\left( \lim_{L\rightarrow \infty }\{\mathrm{h}_{L}^{\mathfrak{m}},f\}\right)
\notag \\
&=&\lim_{L\rightarrow \infty }\left\{ \mathrm{h}_{L}^{\mathfrak{m}},V_{t}^{%
\mathfrak{m}}(f)\right\} ,  \label{Liouville's equations}
\end{eqnarray}%
where all limits have to be understood point-wise on the \emph{dense}
subspace $E_{\mathrm{p}}\subseteq E^{+}$ of all periodic states. We thus
obtain the usual (autonomous) dynamics of classical mechanics written in
terms of Poisson brackets. See, e.g., \cite[Proposition 10.2.3]%
{classical-dynamics}. This corresponds to \emph{Liouville's equation}.

By \cite[p. 34, e.g., Eq. (114)]{BruPedra-MFII}, observe additionally that,
for any $\mathfrak{m}\in \mathcal{M}_{1}^{\infty }$ and $\vec{\ell}\in 
\mathbb{N}^{d}$, the flow preserves the sets $E^{+}$, $E_{\vec{\ell}}$ and $%
\mathcal{E}_{\vec{\ell}}$, i.e.,%
\begin{equation}
\begin{array}{l}
\bigcup\limits_{t\in \mathbb{R}}\mathbf{\varpi }^{\mathfrak{m}}\left(
t;E^{+}\right) \subseteq E^{+} \\ 
\bigcup\limits_{t\in \mathbb{R}}\mathbf{\varpi }^{\mathfrak{m}}\left( t;E_{%
\vec{\ell}}\right) \subseteq E_{\vec{\ell}} \\ 
\bigcup\limits_{t\in \mathbb{R}}\mathbf{\varpi }^{\mathfrak{m}}\left( t;%
\mathcal{E}_{\vec{\ell}}\right) \subseteq \mathcal{E}_{\vec{\ell}}.%
\end{array}
\label{conservation}
\end{equation}%
Therefore, $V_{t,s}^{\mathfrak{m}}$ can be seen as a mapping from $C(E^{+};%
\mathbb{C})$, $C(E_{\vec{\ell}};\mathbb{C})$ or $C(\mathcal{E}_{\vec{\ell}};%
\mathbb{C})$ to itself:%
\begin{equation}
\begin{array}{l}
V_{t}^{\mathfrak{m}}(f|_{E^{+}}):=(V_{t}^{\mathfrak{m}}f)|_{E^{+}} \\ 
V_{t}^{\mathfrak{m}}(f|_{E_{\vec{\ell}}}):=(V_{t}^{\mathfrak{m}}f)|_{E_{\vec{%
\ell}}} \\ 
V_{t}^{\mathfrak{m}}(f|_{\mathcal{E}_{\vec{\ell}}}):=(V_{t}^{\mathfrak{m}%
}f)|_{\mathcal{E}_{\vec{\ell}}}%
\end{array}
\label{translation invaration encore}
\end{equation}%
for any $t\in \mathbb{R}$, $f\in C(E;\mathbb{C})$, $\mathfrak{m}\in \mathcal{%
M}_{1}^{\infty }$ and $\vec{\ell}\in \mathbb{N}^{d}$. By using the Poisson
brackets (\ref{poisson bracketsbis}), Liouville's equation (\ref{Liouville's
equations}) can be written on $C(E^{+};\mathbb{C})$, $C(E_{\vec{\ell}};%
\mathbb{C})$ or $C(\mathcal{E}_{\vec{\ell}};\mathbb{C})$ for any $\mathfrak{m%
}\in \mathcal{M}_{1}^{\infty }$ and $\vec{\ell}\in \mathbb{N}^{d}$.

\begin{remark}
\mbox{ }\newline
The mathematically rigorous derivation of Liouville's equation (\ref%
{Liouville's equations}) is non-trivial and results from Lieb-Robinson
bounds for multi-commutators \cite{brupedraLR}, first derived in 2017.
\end{remark}

\subsubsection{Entanglement of Quantum and Classical Dynamics\label%
{Entanglement}}

In the thermodynamic limit, the \textquotedblleft
primordial\textquotedblright\ algebra is the separable unital $C^{\ast }$%
-algebra $\mathcal{U}$, generated by fermionic annihilation and creation
operators satisfying the canonical anti-commutation relations, as explained
in Appendix \ref{Algebra of Lattices}. Fix once and for all $\mathfrak{m}\in 
\mathcal{M}_{1}^{\infty }$. Let $K=E$, $E^{+}$ or $E_{\vec{\ell}}=\overline{%
\mathcal{E}_{\vec{\ell}}}$ for some $\vec{\ell}\in \mathbb{N}^{d}$, which
is, in each case, a metrizable, convex (weak$^{\ast }$-) compact subset of
the dual space $\mathcal{U}^{\ast }$. \bigskip

\noindent \textbf{(i) Classical dynamics.} The classical (i.e., commutative)
unital $C^{\ast }$-algebra is the algebra $C\left( K;\mathbb{C}\right) $ of
continuous and complex-valued functions on $K$. The mapping $\mathbf{\varpi }%
^{\mathfrak{m}}$, the solution to the self-consistency equation\emph{\ }(\ref%
{self-consistency equation}), yields a strongly continuous group $(V_{t}^{%
\mathfrak{m}})_{t\in \mathbb{R}}$ of $\ast $-automorphisms of $C\left( K;%
\mathbb{C}\right) $, satisfying Liouville's equation as previously
explained.\bigskip

\noindent \textbf{(ii) Quantum dynamics.} Similar to quantum-classical
hybrid theories of theoretical physics, described for instance in \cite%
{extra-ref1988,extra-ref,extra-ref2,extra-ref3,extra-ref00,extra-ref1},
consider now a secondary quantum algebra $C(K;\mathbb{C})\otimes \mathcal{U}$%
, which is nothing else (up to isomorphism) than the $C^{\ast }$-algebra $%
C(K,\mathcal{U})$ of all (weak$^{\ast }$) continuous $\mathcal{U}$-valued
functions on states. By \cite[Proposition 6.2]{BruPedra-MFII} and (\ref%
{conservation}), the mapping $\mathbf{\varpi }^{\mathfrak{m}}$ from $\mathbb{%
R}$ to the space of automorphisms (or self-homeomorphisms) of $K$ leads to a
(state-dependent) quantum dynamics $\mathfrak{T}^{\mathfrak{m}}:=(\mathfrak{T%
}_{t}^{\mathfrak{m}})_{t\in {\mathbb{R}}}$ on 
\begin{equation*}
C\left( K,\mathcal{U}\right) \equiv C\left( K;\mathbb{C}\right) \otimes 
\mathcal{U},
\end{equation*}%
via the strongly continuous,\ state-dependent two-para%
\-%
meter family $(\tau _{t,s}^{\Psi ^{\left( \mathfrak{m},\rho \right)
}})_{s,t\in {\mathbb{R}}}$ with $\Psi ^{\left( \mathfrak{m},\rho \right) }$
defined by (\ref{self-consistency equation2}):%
\begin{equation*}
\left[ \mathfrak{T}_{t}^{\mathfrak{m}}\left( f\right) \right] \left( \rho
\right) :=\tau _{t,0}^{\Psi ^{\left( \mathfrak{m},\rho \right) }}\left(
f\left( \rho \right) \right) \ ,\qquad \rho \in K,
\end{equation*}%
for any function$\ f\in C(K,\mathcal{U})$ and time $t\in \mathbb{R}$.
\bigskip

\noindent \textbf{(iii) Quantum-classical dynamical entanglement.} By
following arguments of \cite[End of Section 5.2]{Bru-pedra-MF-I}, any
(state-dependent) quantum dynamics on $C(K,\mathcal{U})$ letting every
single element of $C(K;\mathbb{C}\mathfrak{1})\subseteq C(K,\mathcal{U})$
invariant yields a classical dynamics, which, in the case of $\mathfrak{T}^{%
\mathfrak{m}}$, is exactly $(V_{t}^{\mathfrak{m}})_{t\in \mathbb{R}}$. More
interestingly, as we remark in \cite[Section 4.2]{Bru-pedra-MF-I}, each 
\textit{classical} Hamiltonian, i.e., a continuously differentiable function
of $C(K;\mathbb{R})$, leads to a state-dependent quantum dynamics. If the
classical Hamiltonian equals (\ref{classical energy}) then the limit quantum
dynamics, when $L\rightarrow \infty $, is precisely $\mathfrak{T}^{\mathfrak{%
m}}$. In other words, on can recover the classical dynamics from the quantum
one, and vice versa. The classical and quantum systems are completely
interdependent, i.e., \emph{entangled}. This view point is very different
from the common understanding\footnote{%
At least in many textbooks on quantum mechanics. See for instance \cite[%
Section 12.4.2, end of the 4th paragraph of page 178]{quantum theory}.} of
the relation between quantum and classical mechanics, which is widely seen
as a limiting case of quantum mechanics, even if there exist physical
features (such as the spin of quantum particles) which do not have a clear
classical counterpart.\bigskip

The physical relevance of the mathematical framework we present here comes
from the fact that it is able to encode the infinite volume dynamics of very
general mean-field models, for initial states which are only required to be
periodic in space. In fact, the classical part of the mean-field dynamics
explicitly appears in the time evolution of extreme periodic states in (\ref%
{long-range dyn}), while the quantum part corresponds to the last integral
over extreme states of (\ref{long-range dyn}). The fact that the initial
state must be a periodic state does not represent a serious constraint since
any initial even state $\rho $ can be approximated by a periodic state
constructed\footnote{%
This is possible because of \cite[Theorem 11.2]{Araki-Moriya}.} from its
restriction $\rho |_{\mathcal{U}_{\Lambda _{l}}}$ to $\mathcal{U}_{\Lambda
_{l}}$ for sufficiently large $l\in \mathbb{N}_{0}$. See, e.g., \cite[Proof
of Proposition 2.3]{BruPedra-MFII}. Since $l\in \mathbb{N}_{0}$ is
arbitrarily large, hence there is no real physical restriction in assuming
that the initial state is a periodic one, noting that the physical states of
fermion systems are always even\footnote{%
If the initial state is not even, we cannot a priori construct a periodic
state from its restriction $\rho |_{\mathcal{U}_{\Lambda }}$ for any $%
\Lambda \in \mathcal{P}_{f}$.}.

\subsection{Permutation-Invariant Lattice Fermi Systems\label{sect
Permutation-Invariant Long-Range Models}\label{sect Permutation-Invariant
Long-Range Models copy(1)}}

\subsubsection{Permutation-Invariant Mean-Field Models\label%
{Permutation-Invariant Long-Range Models}}

Recall that $\mathcal{W}_{\Pi }:=\mathcal{W}^{0}$ is the space of
permutation-invariant (or on-site) interactions, defined by Equation (\ref%
{finite range}) for $\mathrm{R}=0$. Define 
\begin{equation}
\mathcal{M}_{\Pi }:=\left( \mathcal{W}^{\mathbb{R}}\cap \mathcal{W}_{\Pi
}\right) \times \mathcal{S}^{0}.  \label{model permutation invariant}
\end{equation}%
We name it the space of permutation-invariant mean-field models, because all
associated local Hamiltonians are invariant under permutations: Let $\Pi $
be the set of all bijective mappings from $\mathbb{Z}^{d}$ to itself which
leave all but finitely many elements invariant. It is a group with respect
to the composition of mappings. The condition 
\begin{equation}
\mathfrak{p}_{\pi }:a_{x,\mathrm{s}}\mapsto a_{\pi (x),\mathrm{s}},\quad
x\in \mathbb{Z}^{d},\;\mathrm{s}\in \mathrm{S},
\label{definition perm automorphism}
\end{equation}%
defines a group homomorphism $\pi \mapsto \mathfrak{p}_{\pi }$ from $\Pi $
to the group of $\ast $-automorphisms of the $C^{\ast }$-algebra $\mathcal{U}
$. Then, for any $\mathfrak{m}\in \mathcal{M}_{\Pi }$ and $L\in \mathbb{N}%
_{0}$, the local Hamiltonian $U_{L}^{\mathfrak{m}}$ defined by (\ref%
{equation long range energy}) is permutation-invariant, that is, 
\begin{equation}
\mathfrak{p}_{\pi }\left( U_{L}^{\mathfrak{m}}\right) =U_{L}^{\mathfrak{m}%
},\qquad \pi \in \Pi ,\ \pi \left( \Lambda _{L}\right) =\Lambda _{L}.
\label{permutatino invariant hamiltonian}
\end{equation}

An example of permutation-invariant model is given by the strong-coupling
BCS-Hubbard model: Fix $\mathrm{S}=\{\uparrow ,\downarrow \}$. Let $\Phi
^{Hubb},\Psi ^{BCS}\in \mathcal{W}_{\Pi }\cap \mathcal{W}^{\mathbb{R}}$ be
defined by%
\begin{eqnarray*}
&&\Phi _{\left\{ x\right\} }^{Hubb}%
:=%
-\mu \left( n_{x,\uparrow }+n_{x,\downarrow }\right) -h\left( n_{x,\uparrow
}-n_{x,\downarrow }\right) \\
&&\qquad +2\lambda n_{x,\uparrow }n_{x,\downarrow } \\
&&\Psi _{\left\{ x\right\} }^{BCS}%
:=%
a_{x,\downarrow }a_{x,\uparrow }
\end{eqnarray*}%
for $x\in \mathbb{Z}^{d}$ and $\Phi _{\Lambda }^{Hubb}:=0=:\Psi _{\Lambda
}^{BCS}$ otherwise. Let $\mathfrak{a}^{BCS}\in \mathcal{S}^{0}$ be defined,
for all Borel subset $\mathfrak{B}\subseteq \mathbb{S}$, by 
\begin{equation}
\mathfrak{a}^{BCS}\left( \mathfrak{B}\right) :=-\gamma \mathbf{1}\left[ \Psi
^{BCS}\in \mathfrak{B}\right] .  \label{Strong BCS long range}
\end{equation}%
for some $\gamma \geq 0$, with $\mathbf{1}\left[ \cdot \right] $ being the
indicator function\footnote{$\mathbf{1}\left[ p\right] =1$ if the
proposition $p$ holds true and $\mathbf{1}\left[ p\right] =0$ otherwise.}.
Then, 
\begin{equation*}
\mathfrak{m}_{0}:=(\Phi ^{Hubb},\mathfrak{a}^{BCS})\in \mathcal{M}_{\Pi }
\end{equation*}%
is the strong-coupling BCS-Hubbard model since, in this case, the local
Hamiltonian $U_{L}^{\mathfrak{m}_{0}}$ is equal to the strong-coupling
BCS-Hubbard Hamiltonian $\mathrm{H}_{L}$ defined by (\ref{strong coupling
ham}).

\subsubsection{Permutation-Invariant State Space\label{Permutation-Invariant
State Space}}

The set of all permutation-invariant states is defined by%
\begin{equation}
E_{\Pi }:=\{\rho \in E:\rho =\rho \circ \mathfrak{p}_{\pi }\ \text{for}\ 
\text{all}\ \pi \in \Pi \},  \label{permutation inv states}
\end{equation}%
$\mathfrak{p}_{\pi }$ being the unique $\ast $-automorphism of $\mathcal{U}$
satisfying (\ref{definition perm automorphism}). Obviously, 
\begin{equation*}
E_{\Pi }\subseteq \bigcap\limits_{\vec{\ell}\in \mathbb{N}^{d}}E_{\vec{\ell}%
}\subseteq E^{+}.
\end{equation*}%
Furthermore, $E_{\Pi }$ is metrizable, convex and compact and, by \cite[%
Theorem 5.3]{BruPedra2}, for any $\rho \in E_{\Pi }$, there is a unique
probability measure $\mu _{\rho }$ on $E_{\Pi }$ with support in the
(non-empty) set $\mathcal{E}_{\Pi }$ of its extreme points such that%
\begin{equation}
\rho =\int_{\mathcal{E}_{\Pi }}\hat{\rho}\ \mathrm{d}\mu _{\rho }\left( \hat{%
\rho}\right) .  \label{affine decomposition}
\end{equation}%
The set $\mathcal{E}_{\Pi }$ can be characterized by\ a version of the St{\o 
}rmer theorem for permutation-invariant states on the $C^{\ast }$-algebra $%
\mathcal{U}$. This theorem is a non-commutative version of the celebrated de
Finetti theorem of (classical) probability theory. It is proven in the case
of quantum-spin systems in \cite{Stormer} and for the fermion algebra $%
\mathcal{U}$ in \cite[Lemmata 6.6-6.8]{BruPedra1}. It asserts that extreme
permutation-invariant states $\rho \in \mathcal{E}_{\Pi }$ are product
states defined as follows: First recall that the space $E_{\Lambda }^{+}$ of
finite volume even states is defined by (\ref{finite volume even states})
for any $\Lambda \in \mathcal{P}_{f}$. Then, via \cite[Theorem 11.2]%
{Araki-Moriya}, for any $\rho _{0}\in E_{\{0\}}^{+}$, there is a unique even
state 
\begin{equation}
\rho :=\otimes _{\mathbb{Z}^{d}}\rho _{0}\in E^{+}
\label{product state extremes0}
\end{equation}%
satisfying%
\begin{equation}
\rho (\alpha _{x_{1}}(A_{1})\cdots \alpha _{x_{n}}(A_{n}))=\rho
_{0}(A_{1})\cdots \rho _{0}(A_{n})  \label{product states}
\end{equation}%
for all $A_{1}\ldots A_{n}\in \mathcal{U}_{\{0\}}$ and all $x_{1},\ldots
x_{n}\in \mathbb{Z}^{d}$ such that $x_{i}\not=x_{j}$ for $i\not=j$. Recall
that $\alpha _{x}$, $x\in \mathbb{Z}^{d}$, defined by (\ref{transl}), are
the $\ast $-automorphisms of $\mathcal{U}$ that represent translations. The
set of all states of the form (\ref{product state extremes0}), called \emph{%
product states}, is denoted by $E_{\otimes }$. It is nothing else but the
set $\mathcal{E}_{\Pi }$ of extreme points of $E_{\Pi }$, i.e., 
\begin{equation}
E_{\otimes }=\mathcal{E}_{\Pi }.  \label{product state extremes}
\end{equation}%
This identity refers to the St{\o }rmer theorem, see, e.g., \cite[Theorem 5.2%
]{BruPedra2}.

Since product states are particular extreme states\footnote{%
By \cite[Theorem 5.2]{BruPedra2}, all product states are strongly mixing,
which means \cite[Eq. (1.10)]{BruPedra2}. They are, in particular, strongly
clustering and thus ergodic with respect to any sub-groups $(\mathbb{Z}_{%
\vec{\ell}}^{d},+)\subseteq (\mathbb{Z}^{d},+)$, where $\vec{\ell}\in 
\mathbb{N}^{d}$. By \cite[Theorem 1.16]{BruPedra2}, all product states
belong to $\mathcal{E}_{\vec{\ell}}$ for any $\vec{\ell}\in \mathbb{N}^{d}$.}
of $E_{\vec{\ell}}$ for any $\vec{\ell}\in \mathbb{N}^{d}$, it follows from (%
\ref{product state extremes}) that 
\begin{equation}
\mathcal{E}_{\Pi }=E_{\otimes }\subseteq \bigcap\limits_{\vec{\ell}\in 
\mathbb{N}^{d}}\mathcal{E}_{\vec{\ell}}  \label{product state extremes2}
\end{equation}%
and the set $E_{\Pi }\subseteq E_{\vec{\ell}}$ is thus a closed metrizable
face\footnote{%
A face $F$ of a convex set $K$ is defined to be a subset of $K$ with the
property that, if $\rho =\lambda _{1}\rho _{1}+\cdots +\lambda _{n}\rho
_{n}\in F$ with $\rho _{1},\ldots ,\rho _{n}\in K$, $\lambda _{1},\ldots
,\lambda _{n}\in (0,1)$ and $\lambda _{1}+\cdots +\lambda _{n}=1$, then $%
\rho _{1},\ldots ,\rho _{n}\in F$.} of $E_{\vec{\ell}}$. For a more thorough
exposition on this subject, see \cite[Section 5.1]{BruPedra2}. By (\ref%
{product state extremes}), the extreme boundary $\mathcal{E}_{\Pi }$ of $%
E_{\Pi }$ is also closed and, in contrast with $E$, $E^{+}$ and $E_{\vec{\ell%
}}$ for any $\vec{\ell}\in \mathbb{N}^{d}$, $\mathcal{E}_{\Pi }$ is \emph{not%
} a dense subset of $E_{\Pi }$. This is not surprising since states of $%
\mathcal{E}_{\Pi }=E_{\otimes }$ are in one-to-one correspondence with even
states on the finite-dimensional $C^{\ast }$-algebra $\mathcal{U}_{\{0\}}$.

\subsubsection{Quantum Part of Permutation-Invariant Mean-Field Dynamics 
\label{Quantum Part of Permutation-Invariant Long-Range Dynamics}}

Fix once and for all $\mathfrak{m}\in \mathcal{M}_{\Pi }$. If $\rho \in
E_{1}:=E_{(1,\cdots ,1)}$, i.e., it is translation-invariant, then the
approximating interaction (\ref{approx interaction}) satisfies 
\begin{equation}
\Phi ^{(\mathfrak{m},\rho )}=\Phi ^{(\mathfrak{m},\rho |_{\mathcal{U}%
_{\{0\}}})}\in \mathcal{W}_{\Pi }\cap \mathcal{W}^{\mathbb{R}}
\label{state sfsdf}
\end{equation}%
and the infinite volume dynamics constructed from this interaction, as
defined by (\ref{dynamics limit}), preserves the local $C^{\ast }$-algebra $%
\mathcal{U}_{\Lambda }$ for any $\Lambda \in \mathcal{P}_{f}$. By (\ref%
{cauchy2})-(\ref{dynamics limit}) and (\ref{self-consistency equation})-(\ref%
{self-consistency equation2}), it also follows that%
\begin{equation}
\begin{array}{c}
\bigcup\limits_{t\in \mathbb{R}}\mathbf{\varpi }^{\mathfrak{m}}\left(
t;E_{\Pi }\right) \subseteq E_{\Pi }\subseteq E_{1} \\ 
\bigcup\limits_{t\in \mathbb{R}}\mathbf{\varpi }^{\mathfrak{m}}\left(
t;E_{\otimes }\right) \subseteq E_{\otimes }\subseteq E_{\Pi }%
\end{array}
\label{conservation2}
\end{equation}%
(compare with (\ref{conservation})) and, for any $\Lambda \in \mathcal{P}%
_{f} $, $t\in \mathbb{R}$ and translation-invariant state $\rho \in
E_{1}\supseteq E_{\Pi }$, 
\begin{equation}
\mathbf{\varpi }^{\mathfrak{m}}\left( t;\rho \right) |_{\mathcal{U}_{\Lambda
}}=\mathbf{\varpi }^{\mathfrak{m}}\left( t;\rho |_{\mathcal{U}_{\Lambda
}}\right) |_{\mathcal{U}_{\Lambda }}\in E_{\Lambda }^{+}
\label{eq restricted0}
\end{equation}%
with $E_{\Lambda }^{+}$ being the space of finite volume even states defined
by (\ref{finite volume even states}) for any $\Lambda \in \mathcal{P}_{f}$.

If the initial state $\rho \in E_{\Pi }$ is permutation-invariant, then, by (%
\ref{long-range dyn}), (\ref{affine decomposition}) and (\ref{product state
extremes2}), there is a unique probability measure $\mu _{\rho }$ on $E_{\Pi
}$ with support in $\mathcal{E}_{\Pi }=E_{\otimes }$ such that, for any $%
A\in \mathcal{U}$, 
\begin{eqnarray}
&&\lim_{L\rightarrow \infty }\rho \circ \tau _{t}^{(L,\mathfrak{m})}\left(
A\right)  \notag \\
&=&\int_{E_{\otimes }}\mathbf{\varpi }^{\mathfrak{m}}\left( t;\hat{\rho}%
\right) \left( A\right) \ \mathrm{d}\mu _{\rho }\left( \hat{\rho}\right) 
\notag \\
&=&\int_{E_{\otimes }}\hat{\rho}\circ \tau _{t,0}^{\Psi ^{\left( \mathfrak{m}%
,\hat{\rho}\right) }}\left( A\right) \ \mathrm{d}\mu _{\rho }\left( \hat{\rho%
}\right)  \label{sdssd}
\end{eqnarray}%
with $\mathbf{\varpi }^{\mathfrak{m}}$ being the solution to the
self-consistency equation\emph{\ }(\ref{self-consistency equation}). In
particular, by (\ref{conservation2}), the time-evolution of a
permutation-invariant state is uniquely determined by its restriction to the
finite-dimen%
\-%
sional subalgebra $\mathcal{U}_{\{0\}}$ (dimension $2^{2\left\vert \mathrm{S}%
\right\vert }$).

If the initial state $\rho \in E_{1}\supseteq E_{\Pi }$ is
translation-invariant, then Equation (\ref{long-range dyn}) restricted to
the finite-dimensional $C^{\ast }$-algebra $\mathcal{U}_{\Lambda }$ with $%
\Lambda \in \mathcal{P}_{f}$ reads\footnote{%
Note that $\mu _{\rho }$ in (\ref{long-range dyn}) is a probability measure
on $E_{1}\subseteq E^{+}$, but since the restriction mapping $\rho \mapsto
\rho |_{\mathcal{U}_{\Lambda }}$ is continuous for any $\Lambda \in \mathcal{%
P}_{f}$, $\mu _{\rho }$ can be pushed forward to a probability measure on $%
E_{\Lambda }^{+}$, which we also denote $\mu _{\rho }$.}%
\begin{equation}
\lim_{L\rightarrow \infty }\rho |_{\mathcal{U}_{\Lambda }}\circ \tau
_{t}^{(L,\mathfrak{m})}\left( A\right) =\int_{E_{\Lambda }^{+}}\mathbf{%
\varpi }^{\mathfrak{m}}\left( t;\hat{\rho}\right) \left( A\right) \ \mathrm{d%
}\mu _{\rho }\left( \hat{\rho}\right)  \label{eq restricted}
\end{equation}%
for any $A\in \mathcal{U}_{\Lambda }$. For each fixed $\Lambda \in \mathcal{P%
}_{f}$, this gives now a family of equations on the finite-dimensional
algebra $\mathcal{U}_{\Lambda }$ (dimension $2^{2\left\vert \Lambda
\right\vert \times \left\vert \mathrm{S}\right\vert }$). These equations
completely determine the time-evolution of a translation-invariant initial
states.

For any $\vec{\ell}$-periodic state $\rho \in E_{\vec{\ell}}$ ($\vec{\ell}%
\in \mathbb{N}^{d}$), the approximating interaction (\ref{approx interaction}%
) also belongs to $\mathcal{W}_{\Pi }\cap \mathcal{W}^{\mathbb{R}}$. The
only difference with respect to translation-invariant states is that the
on-site state $\rho |_{\mathcal{U}_{\{0\}}}$ in (\ref{state sfsdf}) has to
be replaced with the finite volume state $\rho |_{\mathcal{U}_{\mathcal{Z}_{%
\vec{\ell}}}}$, where, for $\vec{\ell}=(\ell _{1},\ldots ,\ell _{d})\in 
\mathbb{N}^{d}$, 
\begin{equation*}
\mathcal{Z}_{\vec{\ell}}:=\left\{ (x_{1},\ldots ,x_{d})\in \mathbb{Z}%
^{d}:x_{i}\in \{0,\ldots ,\ell _{i}-1\}\right\} .
\end{equation*}%
Compare, as an example, with (\ref{ddd}). Hence, if the initial state is
periodic then Equation (\ref{long-range dyn}) leads again to a family of
equations on the finite-dimensional algebra $\mathcal{U}_{\Lambda }$
(dimension $2^{2\left\vert \Lambda \right\vert \times \left\vert \mathrm{S}%
\right\vert }$) for each $\Lambda \in \mathcal{P}_{f}$ such that\footnote{%
The restriction $\Lambda \supseteq \mathcal{Z}_{\vec{\ell}}$ can also be
easily understood by seeing $\vec{\ell}$-periodic states as a
translation-invariant state on the CAR $C^{\ast }$-algebra with new spin set 
$\mathcal{Z}_{\vec{\ell}}\times \mathrm{S}$.} $\Lambda \supseteq \mathcal{Z}%
_{\vec{\ell}}$. These equations again determine the time-evolution of a
periodic initial state.

\subsubsection{Classical Part of Permutation-Invariant Mean-Field Dynamics 
\label{Classical Part of Permutation-Invariant Long-Range Dynamics}}

Fix again once and for all $\mathfrak{m}\in \mathcal{M}_{\Pi }$. By (\ref%
{conservation2}), the strongly continuous group $(V_{t}^{\mathfrak{m}%
})_{t\in \mathbb{R}}$ of $\ast $-auto%
\-%
morphisms defined by (\ref{classical evolution family}) can be restricted to
the unital $C^{\ast }$-algebra $C(E_{\otimes };\mathbb{C})$ of continuous
functions on the compact space $E_{\otimes }$ of product states. See also 
\cite[Section 5.4 with $\mathcal{B}=\mathcal{U}_{\{0\}}$]{Bru-pedra-MF-I}.
Without any risk of confusion, we denote the restriction of $(V_{t}^{%
\mathfrak{m}})_{t\in \mathbb{R}}$ to $E_{\otimes }$ again by $(V_{t}^{%
\mathfrak{m}})_{t\in \mathbb{R}}$.

Using (\ref{product state extremes0})-(\ref{product state extremes}) we
identify $E_{\otimes }$ with the space $E_{\{0\}}^{+}$ of on-site even
states and see now $(V_{t}^{\mathfrak{m}})_{t\in \mathbb{R}}$ as acting on
the algebra $C(E_{\{0\}}^{+};\mathbb{C})$. Similar to (\ref{polynomials}),
the set of polynomials in this space of functions is denoted by 
\begin{equation*}
\mathbb{P}_{\{0\}}:=\mathbb{C}[\{\hat{A}|_{E_{\{0\}}^{+}}:A\in \mathcal{U}%
_{\{0\}}\}]\subseteq C(E_{\{0\}}^{+};\mathbb{C}).
\end{equation*}%
Local classical energy functions \cite[Definition 6.9]{BruPedra-MFII} on $%
\mathcal{U}_{\{0\}}$ are defined by $\mathrm{h}_{0}^{\mathfrak{m}%
}|_{E_{\{0\}}^{+}}$, where, by (\ref{local energy}) and (\ref{classical
energy}), 
\begin{multline*}
\mathrm{h}_{0}^{\mathfrak{m}}=\widehat{\Phi _{\{0\}}}+\sum_{n\in \mathbb{N}%
}\int_{\mathbb{S}^{n}}\widehat{\Psi _{\{0\}}^{(1)}}\cdots \widehat{\Psi
_{\{0\}}^{(n)}} \\
\mathfrak{a}_{n}\left( \mathrm{d}\Psi ^{(1)},\ldots ,\mathrm{d}\Psi
^{(n)}\right) .
\end{multline*}%
Then, for any time $t\in \mathbb{R}$ and polynomials $f\in \mathbb{P}%
_{\{0\}} $, Liouville's equation (\ref{Liouville's equations}) restricted to
the algebra $C(E_{\{0\}}^{+};\mathbb{C})$ equals 
\begin{equation}
\partial _{t}V_{t}^{\mathfrak{m}}\left( f\right) =V_{t}^{\mathfrak{m}}\left(
\{\mathrm{h}_{0}^{\mathfrak{m}},f\}\right) =\left\{ \mathrm{h}_{0}^{%
\mathfrak{m}},V_{t}^{\mathfrak{m}}(f)\right\} ,
\label{Liouville's equations permutation}
\end{equation}%
where, for any $n,m\in \mathbb{N}$, $A_{1},\ldots ,A_{n}\in \mathcal{U}$, $%
B_{1},\ldots ,B_{m}\in \mathcal{U}$, $g\in C^{1}\left( \mathbb{R}^{n},%
\mathbb{C}\right) $ and $h\in C^{1}\left( \mathbb{R}^{m},\mathbb{C}\right) $%
, 
\begin{equation}
\{\Gamma _{h}|_{\mathcal{U}_{\{0\}}},\Gamma _{g}|_{\mathcal{U}%
_{\{0\}}}\}:=\{\Gamma _{h},\Gamma _{g}\}|_{\mathcal{U}_{\{0\}}}\in
C(E_{\{0\}}^{+},\mathbb{C)}  \label{poisson2}
\end{equation}%
defines again a Poisson bracket, which can be extended to the space $%
C^{1}(E_{\{0\}}^{+};\mathbb{C)}$ of continuously differentiable functions.
Similar to (\ref{sdssd}), Liouville's equation (\ref{Liouville's equations
permutation}) is now written on the finite-dimensional algebra $\mathcal{U}%
_{\{0\}}$ (dimension $2^{2\left\vert \mathrm{S}\right\vert }$) and
completely determines a continuous flow on the compact space $E_{\otimes }$
of product states.\bigskip

\noindent \textit{Acknowledgments:} This work is supported by CNPq
(309723/2020-5), FAPESP (2017/22340-9), as well as by the Basque Government
through the grant IT641-13 and the BERC 2018-2021 program, and by the
Spanish Ministry of Science, Innovation and Universities: BCAM Severo Ochoa
accreditation SEV-2017-0718, MTM2017-82160-C2-2-P.

\end{document}